\documentclass[twocolumn]{aastex61}

\usepackage{breakurl}
\usepackage{amsmath,amssymb,xspace}
\usepackage{fp}

\defcitealias{beaton_2016}{Paper~I}
\defcitealias{hatt_2017}{Paper~II}
\defcitealias{jang_2018}{Paper~III}
\defcitealias{hatt_2018a}{Paper~IV}
\defcitealias{hatt_2018b}{Paper~V}
\defcitealias{cchp2proposal}{CCHP}
\defcitealias{Freedman2019}{Paper~VIII}
\defcitealias{Hoyt2019}{Paper~VI}
\defcitealias{janglee_2017}{JL17}
\defcitealias{lee_2012}{LJ12}
\defcitealias{rizzi_2007}{R07}
\defcitealias{shappee_2011}{SS11}
\defcitealias{sakai_2004}{S04}
\defcitealias{Tikhonov_2015}{T15}
 
 \newcommand{\hst}{\emph{HST}\xspace}
 \newcommand{\galex}{\emph{GALEX}\xspace}
 \newcommand{\gaia}{\emph{Gaia}\xspace}
 
 \newcommand{\wise}{\emph{WISE}\xspace}
\newcommand{\sne}{SNe~Ia\xspace}
\newcommand{\sn}{SN~Ia\xspace}
\newcommand{\gal}{M\,101\xspace}
\newcommand{\snfe}{SN\,2011fe\xspace}
\newcommand{\sigmasmooth}{$\sigma_s$\xspace}
\newcommand{\hc}{$H_{0}$\xspace}

\newcommand{\freedmansub}{AAS17486}
\newcommand{\hoytsub}{AAS17233}
\newcommand{\responsetoref}[1]{{\color{black} #1}}
\usepackage{mathtools}
\DeclarePairedDelimiter{\abs}{\lvert}{\rvert}

\FPeval{\apinffoursevenfive}{round(0.100013,3)} 
\FPeval{\apinffivefivefive}{round(0.0964472,3)} 
\FPeval{\apinfsixzerosix}{round(0.0952613,3)}   
\FPeval{\apinfeightonefour}{round(0.0976345,3)} 

\newcommand{\Iextinction}{0.016}
\FPeval{\IextinctioROUNDED}{round(\Iextinction,2)}
\newcommand{\Vextinction}{0.024}
\FPeval{\VextinctioROUNDED}{round(\Vextinction,2)}
\newcommand{\oldtrgblum}{-4.029}
\newcommand{\oldtrgblumstaterr}{0.011} 
\newcommand{\oldtrgblumsyserr}{0.041}
\newcommand{\trgblum}{-4.049}
\newcommand{\trgblumstaterr}{0.022} 
\newcommand{\trgblumsyserr}{0.039}

\FPeval{\Iextinctionerr}{\Iextinction/2}
\FPeval{\IextinctionerrROUNDED}{round(\Iextinction/2,2)}
\FPeval{\IextinctionerrTOTAL}{round((\Iextinctionerr^2+0.01^2)^0.5,2)}
\newcommand{\ZPerr}{0.02}
\newcommand{\EEerr}{0.02}
\newcommand{\Apcorrerr}{0.01}

\newcommand{\trgbobsval}{25.035}
\newcommand{\trgbobsvalstaterr} {0.030} 
\newcommand{\trgbobsvalsyserr} {0.022}

 \FPeval\trgbobsvalROUNDED{round(\trgbobsval,2)}
 \FPeval\trgbredcorrval{round(\trgbobsval-\Iextinction,2)}
 \FPeval\truetrgbdmod{round(\trgbobsval-\Iextinction-\trgblum,2)}
 \FPeval\truetrgbdmodMpc{round(10^(\truetrgbdmod/5)/100000,2)}
 \FPeval\trgbobsvalstaterrROUNDED{round(\trgbobsvalstaterr,2)}
 \FPeval\trgbobsvalsyserrROUNDED{round((\trgbobsvalsyserr^2+\ZPerr^2+\EEerr^2+\Apcorrerr^2)^0.5,2)}
\FPeval\trgbcorrsvalsyserrROUNDED{round( (\trgbobsvalsyserr^2+\ZPerr^2+\EEerr^2+\Apcorrerr^2+(\IextinctionerrTOTAL)^2 )^0.5,2)}

 \FPeval\dmodcombinedstaterr{ round( (\trgbobsvalstaterr^2+\trgblumstaterr^2)^0.5,2) }
 \FPeval\dmodcombinedsyserr{ round( (\trgbobsvalsyserr^2+\trgblumsyserr^2+\ZPerr^2+\EEerr^2+\Apcorrerr^2+\Iextinctionerr^2)^0.5,2) }

 \FPeval\truetrgbdmodMpcupperrdiststat{ 10^( (\truetrgbdmod+\dmodcombinedstaterr) /5)/100000 }
 \FPeval\truetrgbdmodMpclowerdiststat{ 10^( (\truetrgbdmod-\dmodcombinedstaterr) /5)/100000 }
 \FPeval\truetrgbdmodMpcstaterr{ round( 0.5*(\truetrgbdmodMpcupperrdiststat - \truetrgbdmodMpclowerdiststat) ,2) }
 \FPeval\truetrgbdmodMpcupperrdistsys{ 10^( (\truetrgbdmod+\dmodcombinedsyserr) /5)/100000 }
 \FPeval\truetrgbdmodMpclowerdistsys{ 10^( (\truetrgbdmod-\dmodcombinedsyserr) /5)/100000 }
 \FPeval\truetrgbdmodMpcsyserr{ round( 0.5*(\truetrgbdmodMpcupperrdistsys - \truetrgbdmodMpclowerdistsys) ,2) }
 
 \newcommand{\truetrgbdmodwerr}{$\mu_0 = \truetrgbdmod \pm\dmodcombinedstaterr_{stat} \pm\dmodcombinedsyserr_{sys}~\mathrm{mag}$\xspace}
 \newcommand{\truetrgbdmodMpcwerr}{$D = \truetrgbdmodMpc\pm\truetrgbdmodMpcstaterr_{stat}\pm\truetrgbdmodMpcsyserr_{sys}$ Mpc\xspace}
 \newcommand{\oldtrgblumwerr}{$M_{I}^\mathrm{TRGB}=\oldtrgblum\pm\oldtrgblumstaterr_{stat}\pm\oldtrgblumsyserr_{sys}$\xspace}
  \newcommand{\trgblumwerr}{$M_{F814W}^\mathrm{TRGB}=\trgblum\pm\trgblumstaterr_{stat}\pm\trgblumsyserr_{sys}$\xspace}

\newcommand{\ebvtoai}{1.516}
\FPeval{\aic}{round(0.0082*\ebvtoai,2)}
\FPeval{\eaic}{round(0.0082*\ebvtoai/2.,2)}

\FPeval{\aid}{round(0.0082*\ebvtoai,2)}
\FPeval{\eaid}{round(0.0082*\ebvtoai/2.,2)}

\FPeval{\aie}{round(0.0076*\ebvtoai,2)}
\FPeval{\eaie}{round(0.0076*\ebvtoai/2.,2)}

\FPeval{\aif}{round(0.0081*\ebvtoai,2)}
\FPeval{\eaif}{round(0.0081*\ebvtoai/2.,2)}
\newcommand{\tipf}{25.105}
\FPeval{\tipfcorr}{round(\tipf-\aif,2)}
\FPeval{\tipfcorrerr}{round((\trgbobsvalstaterr^2+\eaif^2)^0.5,2)}

\shorttitle{Optical TRGB Distance to \gal}
\shortauthors{Beaton et al.}

\begin{document} 

\title{\textit{The Carnegie-Chicago Hubble Program.} VII. The Distance to \gal via the Optical Tip of the Red Giant Branch Method\footnote{Based on observations made with the NASA/ESA Hubble Space Telescope, obtained at the Space Telescope Science Institute, which is operated by the  Association of Universities for Research in Astronomy, Inc., under NASA contract NAS 5-26555. These observations are associated with programs GO13691, GO14166, GO13737, and GO13364.}}

\correspondingauthor{Rachael L. Beaton}
\email{rbeaton@princeton.edu}

\author[0000-0002-1691-8217]{Rachael L. Beaton}
\altaffiliation{Hubble Fellow}
\altaffiliation{Carnegie-Princeton Fellow}
\affiliation{Department of Astrophysical Sciences, Princeton University, 4 Ivy Lane, Princeton, NJ~08544}

\author[0000-0002-1143-5515]{Mark Seibert} 
\affil{The Observatories of the Carnegie Institution for Science, 813 Santa Barbara St., Pasadena, CA 91101, USA}

\author[0000-0003-2767-2379]{Dylan Hatt} 
\affil{Department of Astronomy \& Astrophysics, University of Chicago, 5640 South Ellis Avenue, Chicago, IL 60637, USA}


\author{Wendy~L.~Freedman} 
\affil{Department of Astronomy \& Astrophysics, University of Chicago, 5640 South Ellis Avenue, Chicago, IL 60637, USA}

\author{Taylor~J.~Hoyt}\affil{Department of Astronomy \& Astrophysics, University of Chicago, 5640 South Ellis Avenue, Chicago, IL 60637, USA}

\author{In~Sung~Jang} \affil{Leibniz-Institut f$\ddot{u}$r Astrophysik Potsdam (AIP), An der Sternwarte 16, 14482 Potsdam, Germany}

\author{Myung~Gyoon~Lee} \affil{Department of Physics \& Astronomy, Seoul National University, Gwanak-gu, Seoul 151-742, Korea}

\author{Barry~F.~Madore}\affil{The Observatories of the Carnegie Institution for Science, 813 Santa Barbara St., Pasadena, CA 91101, USA}\affil{Department of Astronomy \& Astrophysics, University of Chicago, 5640 South Ellis Avenue, Chicago, IL 60637, USA}

\author{Andrew~J.~Monson} \affil{Department of Astronomy \& Astrophysics, The Pennsylvania State University, 525 Davey Lab, University Park, PA 16802, USA}

\author{Jillian~R.~Neeley} \affil{Department of Physics, Florida Atlantic University, 777 Glades Rd, Boca Raton, FL 33431}

\author{Jeffrey~A.~Rich} \affil{The Observatories of the Carnegie Institution for Science, 813 Santa Barbara St., Pasadena, CA 91101, USA}

\author{Victoria~Scowcroft} \affil{Department of Physics, University of Bath, Claverton Down, Bath, BA2 7AY, United Kingdom} \affil{50th Anniversary Prize Fellow}

\begin{abstract} 

The \emph{Carnegie-Chicago Hubble Program} (CCHP) is building a direct path to the Hubble constant (\hc) using Population II stars as the calibrator of the \sn-based distance scale. 
\responsetoref{This path to calibrate the \sne is independent of the systematics in the traditional Cepheid-based technique.}
In this paper, we present the distance to \gal, the host to \snfe, using the $I$-band tip of the red giant branch (TRGB) based on observations from the ACS/WFC instrument on the \emph{Hubble Space Telescope}.  
The CCHP targets the halo of \gal where there is little to no host-galaxy dust, the red giant branch is isolated from nearly all other stellar populations, and there is virtually no source confusion or crowding at the magnitude of the tip.
Applying the standard procedure for the TRGB method from the other works in the CCHP series, we find an foreground-extinction-corrected \gal distance modulus of \truetrgbdmodwerr, which corresponds to a distance of \truetrgbdmodMpcwerr. 
This result is \responsetoref{consistent} with several recent Cepheid-based determinations, suggesting agreement between Population I and II distance scales for this nearby \sn-host galaxy.
We further analyze four archival datasets for \gal that have targeted its outer disk to argue that targeting in the stellar halo provides much more reliable distance measurements from the TRGB method \responsetoref{due to the combination of multiple structural components and heavily population contamination. Application of the TRGB in complex regions will have sources of uncertainty not accounted for in commonly used uncertainty measurement techniques.} 
\end{abstract}


 \section{Introduction} \label{sec:intro} 

\gal holds an important place in the history of the extragalactic distance scale.
As one of the few nearby, large, and face-on spiral galaxies, \gal was a natural stage for the testing of various distance measurement techniques. 
Cepheids have long been the standard tool for the extragalactic distance scale, but early attempts by \citet{sandage_1974} were unable to find Cepheids on their photographic plates of \gal taken with the 200'' Palomar telescope. 
From the non-detection of Cepheids to a limit of $B\sim$22.5~mag in \gal, `the minimum modulus is rather $\geq$ 29.0~mag' \citep[][their section IV,a,iii]{sandage_1974}. 
As a result, M\,101 at that time set the limit on distance determination via individual Cepheid stars using photographic plates. 
For at least another decade, all distances in the ladder for objects more distant than \gal were determined via other techniques.\footnote{A representative history of alternative distance estimators for the case of M\,101 is vividly given in the account of \citet{lonely_hearts}.} 

A decade after this initial work, \citet{cook_1986} published the discovery of two Cepheids from imaging acquired using CCD detectors and effectively initiated a new era for the distance ladder ($\mu$=29.38~mag). 
Later work by this group \citep{cook_1989,alves_1995} continued to discover Cepheids using largely ground-based data and produced additional refinements of the distance to \gal. 
As a part of the \hst Key Project \citep{freedman_2001}, \citet{kelson_1996} discovered 29 Cepheids in the disk of \gal with WFPC2 and determined a distance modulus of 29.34 $\pm$ 0.17~mag (D = 7.4 $\pm$ 0.6 Mpc). 
Since that time numerous other works have addressed the distance to \gal using Cepheids and have refined the measurement \citep[see the compilation given in][]{lee_2012}.

Historically, \gal served as a useful proving ground for distance measurement techniques and as a ``rung'' in the extragalactic distance ladder. 
However, with the appearance of \snfe \citep{11fe_discovery,nugent_2011}, it has become a vital target for measuring the Hubble constant (\hc) via the modern streamlined route, e.g. that using Cepheids to anchor the \sne zero point.
Not only is \gal home to the most nearby \sn observed with modern CCD detectors, but \snfe also resides in a region of low host-galaxy extinction and has come to be a powerful probe of \sne physics \citep[e.g., see][and references therein]{shappee_2016}.
Thus, a precise distance to \gal is an important component of the modern extragalactic distance scale.

Although Cepheids have now been discovered in more distant \sn hosts than \gal, the difficulty in accurately measuring them---whether due to unknown levels of extinction in dusty spiral arms, inaccurate sky-background estimates due to crowding, or limited observations because of the high cost of multi-epoch imaging, among others---has motivated \sn re-calibration efforts of local galaxies through independent methodologies with arguably fewer systematics. 
The goal of this paper is to provide an independent distance measure to \gal using Population II stars within the context of the \emph{Carnegie-Chicago Hubble Program: (CCHP)} \citep[PI:][G013691]{prop13691}; an overview of this effort is given in \citet[][Paper I]{beaton_2016}.

The CCHP is a multi-facility program aimed at building a measurement of \hc~ \responsetoref{that uses \sne calibrated via a technique that is fully independent of, but parallel to the traditional Cepheid route.}
Our cornerstone standard candle is the Tip of the Red Giant Branch (TRGB), which is the discontinuity in the RGB luminosity function resulting from low-mass stellar evolution; more specifically, at the conclusion of the RGB stellar phase, there is a rapid onset of He-core burning that causes the stars to evolve away from the RGB to the lower-luminosity Horizontal Branch.
Precision TRGB distances hinge on its application to old, metal-poor stars, which can be reliably imaged in the low-extinction, low-crowding stellar halos of galaxies, which, in turn, eliminates or minimizes uncertainties due to internal extinction, point-source crowding, and contamination by other stellar populations. 

The TRGB has entirely different systematics from the Cepheid route and provides not only an independent measurement of \hc, but a means to cross-check Cepheid-based distances. 
In the current era where \hc measured via the traditional distance ladder shows discordance with that from modeling CMB anisotropies \responsetoref{at 4.4-$\sigma$ this independent path is a critical step toward understanding and then reconciling the controversy} \citep{riess_2011,riess_2016,riess_2018,riess_2019,freedman_2010,freedman_2012,freedman_2017}. 

\begin{figure*} 
\centering
\includegraphics[width=\columnwidth]{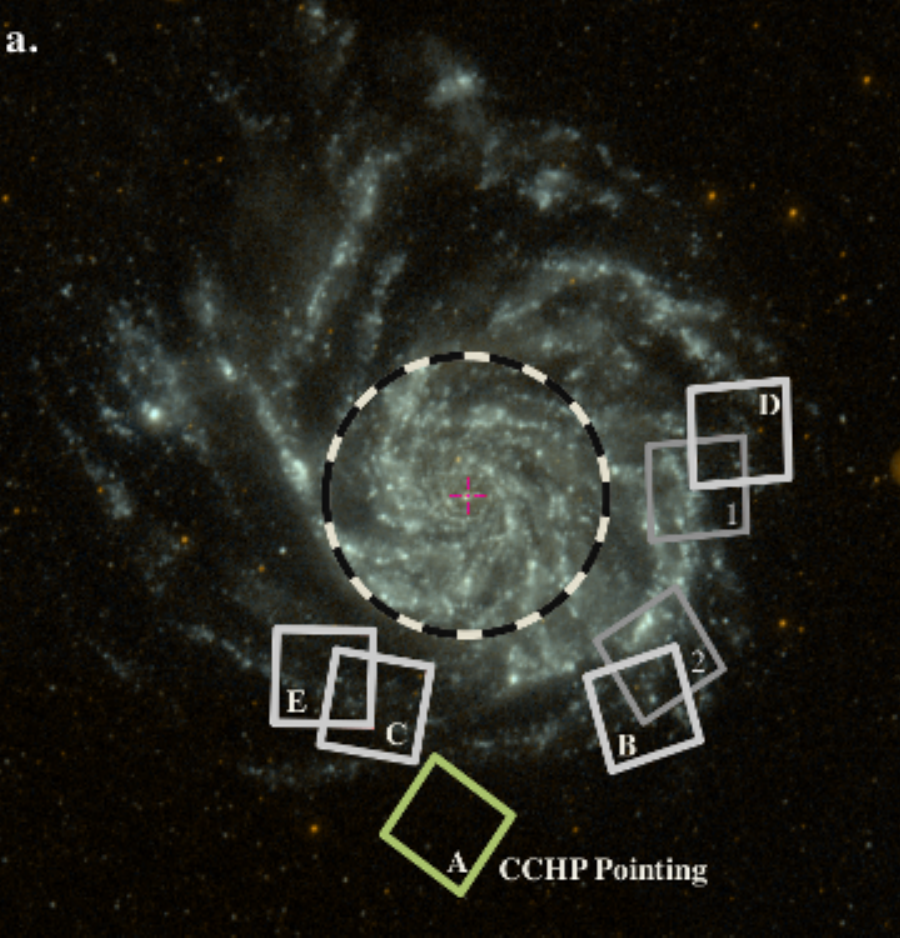}
\includegraphics[width=\columnwidth]{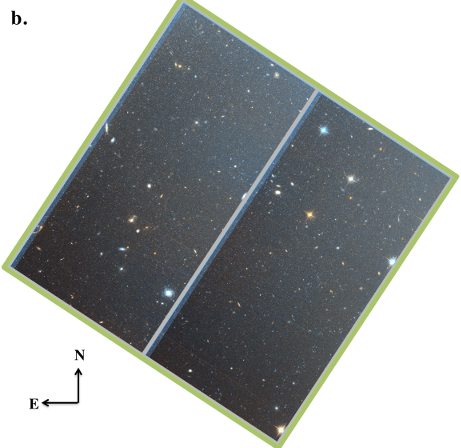}
\caption{ \label{fig:pointingmap}
{\bf (a)} GALEX image of M\,101 with the location of the CCHP Pointing indicated by a green box. The dashed-ring has a radius of 5 arcmin. Other pointings at similar radii are also shown in gray and will be discussed in \autoref{sec:comptrgb}. The CCHP pointing is outside of the UV disk. 
Technical details regarding the pointings are given in \autoref{tab:datasets}.
{\bf (b)} Color image of the CCHP field constructed from co-added F606W and F814W images.}
\end{figure*} 

In \citet[][Paper II]{hatt_2017} we outlined our CCHP reduction techniques for measuring the TRGB, and applied it to the nearby Local Group dwarf, IC\,1613. In \citet[][Paper III]{jang_2018}, we applied and demonstrated the effectiveness of  these techniques to our most distant \sn host galaxy (NGC\,1365). We demonstrated $\sim$2-3\% precision in distance for both galaxies, including detailed discussions and tests of our techniques. Most recently, in \citet[][Paper IV]{hatt_2018a}, we applied this methodology to three \sn host galaxies in the Virgo cluster (NGC\,4424, NGC\,4526 and NGC\,4536); and in \citet[][Paper V]{hatt_2018b}, we measured the distances to NGC\,1316 and NGC\,1448, again with $\sim$3\% precision.
\responsetoref{In \citet[\hoytsub][Paper~VI]{Hoyt2019}, we applied the method to two \sn host galaxies in the Leo Group, again demonstrating the precision and accuracy of the technique.}
In this paper, we now apply these same techniques to \gal, which is the closest \sn host-galaxy in the CCHP.
\responsetoref{\citet[\freedmansub][Paper~VIII]{Freedman2019}}, presents the final TRGB calibration and the value of the \hc determined from the CCHP dataset.

This is not the first Population II distance to \gal.
\responsetoref{Indeed, there are five previous papers that published distance measurements that relied on the TRGB, as well as two distances in the Extragalactic Distance Database\footnote{he EDD is updated routinely and available at this URL: \url{http://edd.ifa.hawaii.edu/dfirst.php}.} \citep[EDD;][]{jacobs_2009}.} 
These distance moduli span a relatively large range, however, 
from $29.05~\pm~0.06$~(stat)~$\pm~0.12$~(sys)~mag \citep{shappee_2011},
  to $29.30~\pm~0.01$~(stat)~$\pm~0.12$~(sys)~mag \citep{lee_2012},
  to $29.34~\pm~0.08$~(stat)~$\pm~0.02$~(sys)~mag \citep{rizzi_2007},
  to $29.42~\pm~0.04$~(stat)~$\pm~0.10$~(sys)~mag \citep{sakai_2004},
 and then most recently a value of 
  $29.15~\pm~0.04$~(stat)~$\pm~0.11$~(sys)~mag \citep{janglee_2017b}. 
In preferred distance from EDD, the is $29.13^{0.08}_{0.09}$~mag.
Thus, while the total quoted uncertainties are of order 0.1~mag for each measurement, the spread in values is four times larger at $\sim$0.4~mag.
At a cursory glance, the TRGB method would seem to be relatively unreliable; though we note the spread for Cepheid distances compiled by \citet{lee_2012} is larger, at 0.7~mag, likely due to the large number of terms in the computation of the distance (to be discussed in \autoref{sec:disc}). 
 
As discussed in \citetalias[][]{beaton_2016}, however, the $\sim$0.4~mag spread in the distance moduli from the TRGB method is most likely due to the use of fields that lie in the disk of \gal, where both crowding from neighboring sources and contamination from intermediate- and young-aged stellar populations are both a large concern for the precise and reliable application of the TRGB method.
These two aspects of the pointing selection can induce strong biases in the measurement of the TRGB.
In this paper, we will demonstrate explicitly the concerns with using disk-dominated fields using a set of four archival datasets.
We place these archival pointings in the context of the CCHP field selection described in \citetalias{beaton_2016} to demonstrate that, at the observation planning stage, one can use surface brightness profiles to bypass the confounding effects of crowding and contamination by selecting an appropriate pointing to ensure a reliable, high precision distance measurement via the TRGB.

The outline of the paper is as follows.
We describe the CCHP data in \autoref{sec:data}.
We determine the apparent magnitude for the TRGB of \gal and its distance from the CCHP data in \autoref{sec:trgb}.
We perform a complementary analysis on literature fields in \autoref{sec:comptrgb}. 
We compare our distance measurements to other techniques in \autoref{sec:disc}.
A summary is given in \autoref{sec:sum}.
Supporting information for the data processing and analyses is provided in \autoref{app}.

\begin{table*} 
\centering
 \caption{\label{tab:datasets} HST+ ACS/WFC Imaging Used in this Work}
\begin{tabular}{ l cc cc c ccc} 
 \hline \hline
Name & $\alpha$$^{a}$ & $\delta$$^{a}$ & \multicolumn{2}{c}{$r_{M101}$}  & Program & \multicolumn{3}{c}{Total Exposure Time (s)} \\
     &  (J2000) & (J2000)  & arcmin & kpc$^{b}$                          &         & $\mathrm{F555W}$ & $\mathrm{F606W}$ & $\mathrm{F814W}$ \\
 \hline \hline 
M101\_A & 14:03:18.800 & +54:09:21.00 & 11.6 & 23.6 & GO13691 (PI: Freedman)            &         & 3750    & 3750   \\
M101\_B & 14:02:30.166 & +54:13:35.58 &  9.6 & 19.5 & GO14166 (PI: Shappee)             &  1884   &         & 1491   \\
M101\_C & 14:03:33.459 & +54:13:24.88 &  8.1 & 16.5 & GO13737 (PI: Shappee)             &  1170   &         & 2138   \\
M101\_D & 14:02:06.461 & +54:23:22.65 &  9.9 & 20.2 & GO13364 (PI: Calzetti)            &         & 1130    & 1420   \\
M101\_E & 14:03:45.960 & +54:14:24.30 &  8.2 & 16.6 & GO13364 (PI: Calzetti)            &         & 1100    & 1400   \\
 \hline \hline
\multicolumn{9}{l}{$^{a}$ Field center may vary between different exposures.} \\
\multicolumn{9}{l}{$^{b}$ Based on NED center (14:03:12.5441, +54:20:56.220) and NED mean distance ($\mu$=29.18~mag or 6.997 Mpc).}
\end{tabular} 
\end{table*} 

\section{Imaging and Photometry} \label{sec:data}

The image data used in this work are described in \autoref{sec:img}.
The analysis of the images to produce photometric catalogs is presented in \autoref{sec:phot}.

\subsection{Imaging Data} \label{sec:img}
We obtained a single \hst+ACS pointing in \gal at a projected radial distance of $R_{\gal}\sim23.6$ kpc (11.6$'$).
The location of this pointing relative to the \gal disk is shown in \autoref{fig:pointingmap}a (green box labelled M101\_A). 
A total exposure time of 3750~s was obtained with three individual images in each of the F606W and F814W filters \citep[G013691;][]{prop13691}. 
\autoref{fig:pointingmap}b shows a color image of the field and shows that there are no strong galactic structures (e.g., spiral arms) in the image; the use of archival GALEX imaging was particularly useful in this regard. 
The pointing was designed following the criteria described in \citetalias{beaton_2016} to minimize contamination from non-RGB stellar populations.
Full details of the pointing are given in \autoref{tab:datasets}. 

In addition to the CCHP field (M101\_A), we also use a series of archival pointings (discussed in detail in \autoref{sec:comptrgb}) that are also indicated in \autoref{fig:pointingmap} with further specifics detailed in \autoref{tab:datasets}.
These data correspond to proposals GO14166 \citep{prop14166}, GO13737 \citep{prop13737}, and GO13364 \citep{prop13364}, which span a range of projected radial separations ranging from 16.5 kpc to 20.2 kpc (8.1$'$ to 9.9$'$). 
The exposure times are $\sim$1/2 to 1/3 of those obtained for the CCHP pointing, which makes them $\sim$1 magnitude shallower. 
Nonetheless, these depths conform to the general signal-to-noise criteria for the CCHP targeting scheme \citepalias{beaton_2016} and are thus ideal for testing the impact of field choice on the precision and accuracy of the TRGB measurements.

\begin{figure} 
\includegraphics[width=\columnwidth]{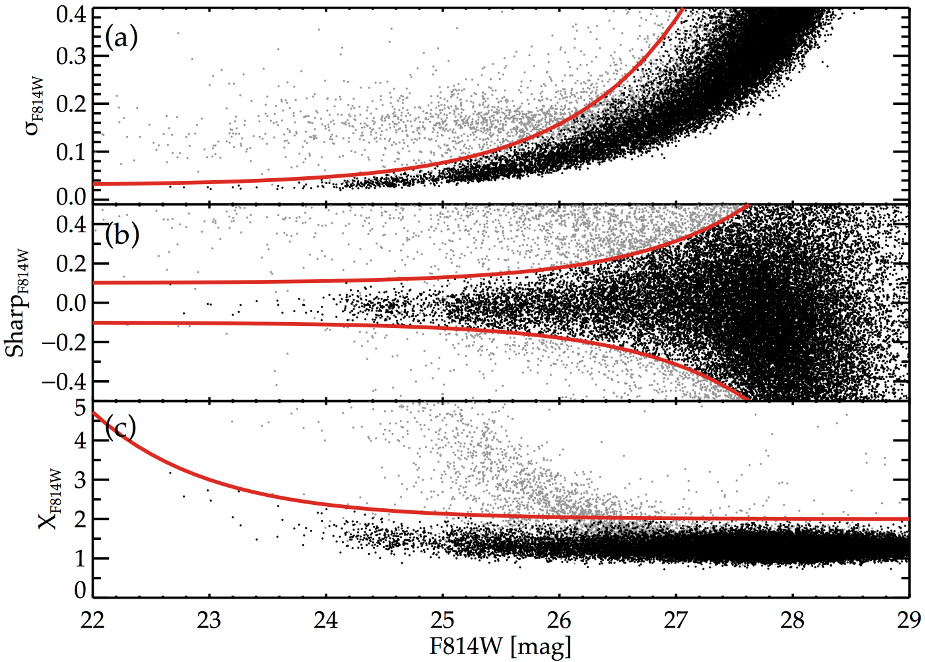}
\centering
\caption{\label{fig:phot_quality} 
Photometric quality for the CCHP field (M101\_A): 
(a) photometric uncertainty in F814W ($\sigma_{F814W}$) as a function of F814W magnitude;
(b) sharp parameter from {\sc DAOPHOT} as a function of F814W magnitude; and, 
(c) chi ($\chi$ parameter as a function of F814W magnitude. 
In each panel, the solid red line is the restriction that was applied to the data to isolate stellar sources, the black points are those sources that pass all three parameter cuts, and the gray points are those sources that fail at least one of the cuts. The strong increase in source density near F814W$\sim$25 mag marks the TRGB and we note that only a handful of sources are marginal at these magnitudes.}
\end{figure} 

\subsection{Photometry}\label{sec:phot}

The photometry for the five fields in \gal were processed identically using an end-to-end pipeline designed to produce homogeneous photometry for the CCHP. 
The pipeline was specifically designed to minimize systematic differences between photometric catalogs of different objects obtained with different integration schemes, while also providing robust measurements of systematics between frames and targets.
This end-to-end approach specifically allows for the investigations to be undertaken both globally in the CCHP and specifically in our comparison of fields in \gal. 

We performed photometry on the STScI processed and charge-transfer-efficiency-corrected individual frames (the FLC data type) retrieved from the Mikulski Archive for Space Telescopes (MAST) for each of the five datasets used in this work (\autoref{tab:datasets}). 
Detailed discussions of the CCHP data processing have been given in \citetalias{hatt_2017} and \citetalias{jang_2018} in application to measurements in IC\,1613 and NGC\,1365, respectively. 
In our previous papers (e.g., \citetalias{hatt_2017}, \citetalias{jang_2018}, \citetalias{hatt_2018a}, and \citetalias{hatt_2018b}), the photometry was implemented in a ``manual'' fashion, whereas the \responsetoref{photometry used in \citetalias[][]{Hoyt2019}, this paper (Paper VII), and \citetalias{Freedman2019} were produced by an end-to-end pipeline.} 
This pipeline is very similar to the manual analyses, but it requires no human intervention and minimizes systematics between fields caused by human choice. 
On the other hand, the pipeline process also eliminates field-by-field optimization as a result of human intervention. 
The general procedures are identical, albeit in some aspects our automated algorithmic approach necessitated both slight adjustments to our aperture core correction procedure and an implementation of sigma-clipping in our mean magnitude computations. 
In the remainder of this section, these changes will be presented in detail within the context of our methodology. 

We used the {\sc daophot} family of programs to perform point-spread-function (PSF) fitting photometry \citep{stetson_1987,stetson_1988,stetson_1990,stetson_1994}.
Instead of fitting an empirical PSF to the data, we create a grid of synthetic PSFs from TinyTim \citep{krist_2011} and use {\sc daophot} to fit an appropriate empirical model.
The same PSF model was used for all frames of the same filter across our program.
A master source list was constructed from an aligned and co-added image built from all of the images, regardless of filter, for a given pointing (using {\sc montage2}). 
This master source list was then used in {\sc allframe} to photometer each of the individual frames simultaneously. 

After the instrumental photometry was computed, we determined an aperture correction for the core of the PSF (ApCore) following the calibration instructions of \citet{sirianni_2005}. 
This was measured by comparing the magnitudes determined from the PSF photometry to the magnitudes in a 0\farcs5 aperture for high signal-to-noise stellar sources in the image. 
Our pipeline implementation of this process differed slightly from that in our previous papers in two ways: (i) The ApCore correction is measured individually for each chip of each frame instead of a single average value computed from an image sequence and (ii) we applied a standardized selection of suitable high signal-to-noise sources instead of a by-eye evaluation. 
Our selection methodology compared the curve-of-growth for a source against the full range of curves of growth from our PSF grid to remove unsuitable sources. 
We removed flux contamination from neighbors using star subtraction routines in {\sc daophot} and retested the profiles.
The ApCore correction was then measured as the uncertainty weighted mean of the sources passing these criteria.
Because we are determining this correction on a frame-by-frame basis, we generally had fewer stars and thereby larger uncertainties on the mean for the ApCore than in previous CCHP works. 
When we visualized the ApCore measurements as a function of image number, we found that the measurements for the two chips move together systematically, which suggests that we were measuring real differences to the PSF during the observation sequence. 
As a result, we think that we are making a better measurement of the ApCore by calculating the correction on a frame-to-frame basis.

After the application of the ApCore for each frame, the mean magnitude for each source was determined using uncertainty weighting and a $\sigma$-clip algorithm (at 2-$\sigma$).
The latter is another feature of the CCHP pipeline that was not implemented in previous reductions that employed an uncertainty-weighting scheme that used all measurements for a given source.
The final {\sc daophot} image quality parameters, $\chi$ and sharp, were then determined as the median value reported for those frames contributing to the mean magnitude for a given source.
We found that the $\sigma$-clipping produces uncertainty and image-quality parameters as a function of magnitude that had the same general shape as those from individual frames whereas those without $\sigma$-clipping often had behavior that does not track individual frame expectations.
A consequence of the $\sigma$-clipping, however, were less well populated catalogs, especially on the faint end, but we are more confident in our ability to assess the photometry across the color-magnitude diagram with this modification.

The instrumental magnitudes were then put onto the STScI photometry system following the description given in \citet{sirianni_2005}.
The specific photometric zeropoints were retrieved on an observation-by-observation basis from the online zeropoint database\footnote{\url{https://acszero-points.stsci.edu/}}, which included updates to the zero points over time \citep[e.g.,][]{mack_2007}.
Following discussion in \citet{sirianni_2005}, we adopted a conservative 2\% systematic uncertainty in flux (0.02~mag) for these zero-points.
The 0\farcs5 to infinite aperture corrections (APInf) were adopted from \citet{boh16} and were converted from the encircled energy (EE) tables into magnitudes. 
Consistent with previous CCHP papers, we adopted a 2\% flux uncertainty in these values \citepalias[0.02~mag; see][for a detailed discussion]{jang_2018}.
We provide the values of these corrections explicitly for each field in \autoref{app} (\autoref{tab:calibration}). 

Our final catalogs were cleaned using a series of photometric quality cuts that are demonstrated in the panels of \autoref{fig:phot_quality}; more specifically,
the photometric uncertainty ($\sigma_{F814W}$) in \autoref{fig:phot_quality}a,
the sharp-parameter (sharp$_{F814W}$) in \autoref{fig:phot_quality}b, and
the chi-parameter ($\chi_{F814W}$) in \autoref{fig:phot_quality}c.
We used a constant$+$exponential function for each of the photometric uncertainty, sharp, and $\chi$ as a function of the F814W magnitude.
This function is shown as the solid red line in each of the panels of \autoref{fig:phot_quality}.
We required that each source pass each of the three cuts, which are shown as the black points of \autoref{fig:phot_quality} with the gray points being those sources that fail one or more of these requirements.
The specific functional form and parameters for each restriction of the photometry are given in \autoref{app}.

\begin{figure} 
\centering
\includegraphics[width=0.9\columnwidth]{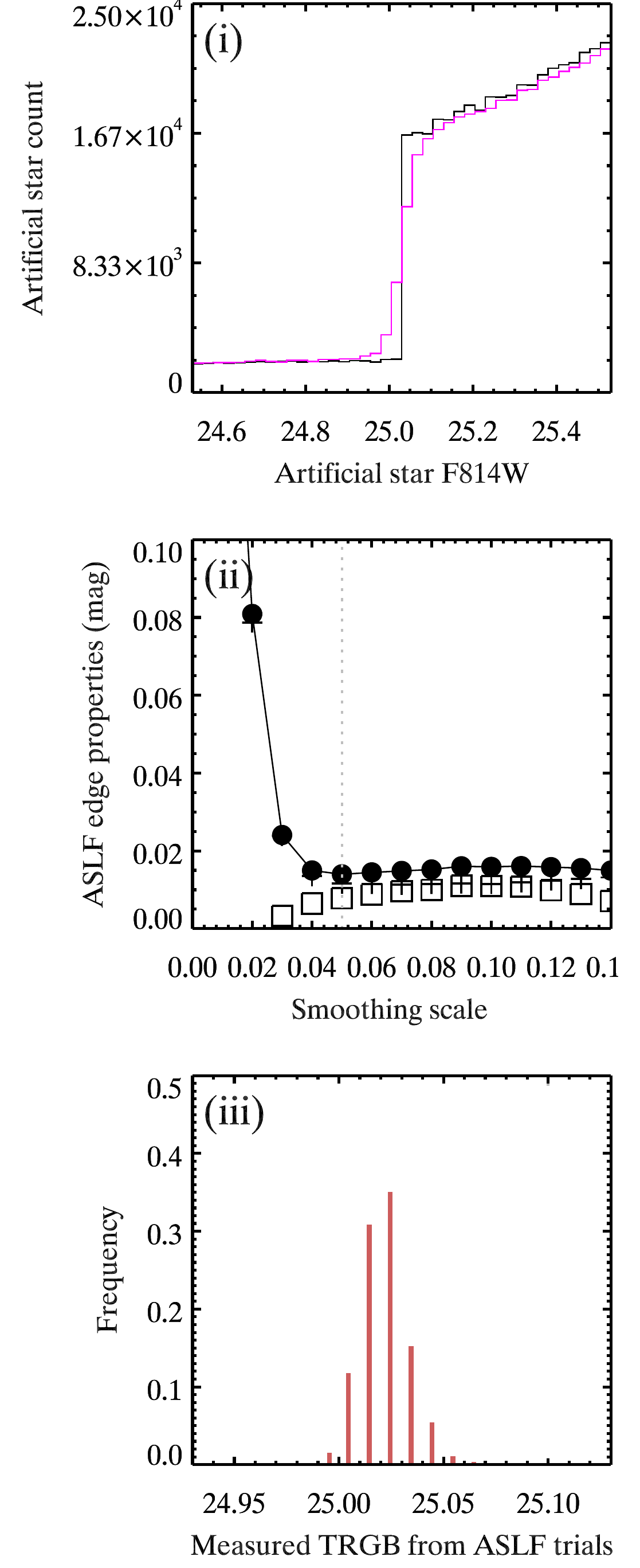}
\caption{\label{fig:aslf_results} 
Uncertainties for our TRGB detection were determined via a series of simulations as described in the text for the CCHP field.
(i) The input (black) and measured (purple) artificial star luminosity functions; 
(ii) The statistical (pluses) and systematic (open boxes) uncertainties and their quadrature sum (filled circles) as a function of \sigmasmooth for simulations of the \gal CCHP field;
(iii) Distribution of measured TRGB values for our selected \sigmasmooth from which we determined the statistical and systematic uncertainties for our TRGB detection. 
}
\end{figure} 

\section{The Tip of the Red Giant Branch} \label{sec:trgb} 

The TRGB is the \responsetoref{truncation} of the RGB sequence due to the lifting of degeneracy in the core of an RGB star when it reaches a specific temperature, corresponding to a critical core mass; the bolometric luminosity coming from the core is thus reasonable approximated by a constant (the He abundance has an effect).
This is both empirically well defined \citep[][among others]{lee_1993,sakai_2004,rizzi_2007} and theoretically supported \citep[][among others]{madore_1999,serenelli_2017}.
Naturally, the energy output from the core is modulated by the composition of the stellar atmosphere as a function of wavelength, which gives rise to the shape of the TRGB in color-magnitude diagrams.
In the optical, higher metallicity corresponds to a progressively steeper, downward-sloping TRGB as incident radiation is ``blanketed'' in the blue and visual bands and then re-emitted thermally in the near-infrared portion of the spectrum. 
The slope trend of the TRGB is consequently reversed when observing in the near-infrared, i.e. higher metallicity stars appear brighter \citep[see][]{hoyt_2018, madore_2018}.
It follows that there is a transition point where the slope of the TRGB as a function of metallicity is approximately flat, or in other words, insensitive to the metal content of the stellar atmosphere.
For stars that are not metal-rich ([Fe/H] \textless -0.5 dex)
the Kron-Cousins $I$ (or similar) bandpass is in this wavelength regime where the TRGB absolute magnitude is constant with color at the few percent level \citep[a recent detailed theoretical exploration is given in][]{serenelli_2017}; these are the types of stars that populate the stellar halo. 
In Pop II systems, therefore, the TRGB provides a remarkably stable standard candle that can be readily identified visually and, as described below, quantified digitally.

Detection of the TRGB discontinuity is typically done by constructing a luminosity function (LF) in the passband of interest by binning the marginalized apparent magnitudes. 
In instances where multiple stellar populations are present, the RGB locus can first be isolated in color-magnitude space using color-cuts. 
As has been discussed in \citetalias{hatt_2017}, \citetalias{jang_2018} \responsetoref{and \citet{beaton_2018}}, there have been many different approaches to analyzing the LF to measure the TRGB.
The two primary approaches are (i) the application an edge detector (to find the point of greatest change in the luminosity function) and (ii) to simultaneously fit LF models of the RGB and other stellar populations that overlap in color-magnitude space.\footnote{These methods typically include include thermally-pulsating asymptotic giant branch stars (TP-AGB) brighter than the TRGB and early-type AGB running parallel to the RGB, although other populations can be present.}
The CCHP has adopted the former approach because it requires fewer assumptions; we do note, however, that it was shown in both \citetalias{hatt_2017} and \citetalias{jang_2018} that a range of TRGB-detection methods appear consistent to within their estimated uncertainties for both the nearby and distant cases of IC\,1613 and NGC\,1365, respectively. 

We bin the F814W magnitudes at 0.01~mag precision to construct a LF for each galaxy and then use the GLOESS smoothing algorithm to reduce the Poisson noise \citep[a description of this algorithm is given in][]{persson_2004,monson_2017,hatt_2017}.
One simple edge detector is the Sobel kernel $[-1,0,+1]$, which is a discrete approximation to the first-derivative. 
We apply this kernel with a signal-to-noise weighting scheme and determine the TRGB magnitude as location of the greatest response in the edge detector (e.g., the point of greatest change in the GLOESS smoothed LF).
The GLOESS algorithm depends critically on the user to input a characteristic kernel width that, in practice, depends on both the quality of the photometry, the number of sources defining the RGB, and the level of contamination from other stellar populations. 
We estimate the optimal value for this smoothing factor (\sigmasmooth) from sets of artificial star experiments; 
we briefly describe this process for \gal and the results are given in \autoref{fig:aslf_results} for the CCHP field.
Motivations for this approach to both determining \sigmasmooth and its associated uncertainties are given in \citetalias{hatt_2017}.

\subsection{Determining \sigmasmooth for the GLOESS Algorithm} \label{sec:aslf}

First, we build an idealized LF with a discontinuity or `jump' in star counts to model the TRGB. 
The LF itself consists of two components: an RGB and an AGB sequence. 
We assume the RGB and AGB LFs have slopes of 0.3~dex and 0.1~dex, respectively.
The input TRGB magnitude itself (i.e. the start of the RGB sequence) is assigned using a preliminary measurement from the real dataset. 
Our RGB LF is designed to extend a full magnitude below the TRGB, while the AGB sequence starts a full magnitude brighter than the TRGB and continues uninterrupted to the bottom of the RGB sequence.

From this input LF, artificial stars are inserted into the CCD images in batches of 2000 stars. 
Colors are assigned to each source from a uniform sampling whose central color is the measured mean of the RGB (approximately $\mathrm{F606W}-\mathrm{F814W}=1.25$~mag) and whose span is the full-color-width of the RGB (approximately $\mathrm{F606W}-\mathrm{F814W}=0.5$~mag). 
The spatial coordinates of the sources ($X$,$Y$) are drawn from a uniform distribution and due to the small number of sources per simulation (2000) these do not strongly change the local crowding at any point in the frame. 
The full photometry process described in \autoref{sec:phot} is performed.
This process is repeated to generate at least 1 million artificial stars, which together make an ``artificial star luminosity function'' (ASLF, hereafter). 
\autoref{fig:aslf_results}(i) gives the input (black) and output (purple) LF for the CCHP field and demonstrates how the sharp input function is rounded by sources of noise in the returned photometry.

To characterize the statistical (random) and systematic uncertainties associated with our TRGB measurement, we model the real CMDs as closely as possible. 
We thus down-sample the ASLF to match the LF-statistics in our frame in a series of trials (e.g., the number of stars in broad bins across the LF). 
For each trial, we apply GLOESS with a single smoothing factor (\sigmasmooth) and apply the Sobel edge-detector.
We repeat this process over a range \sigmasmooth varying from 0.01~mag (no smoothing) to 0.13~mag in 0.01~mag steps for 10,000 trials at each value.
Using the distribution of results for each \sigmasmooth, we determine a statistical uncertainty (the distribution of measurements) and a systematic uncertainty (the mean offset from the input value). 
The results of this process are given in \autoref{fig:aslf_results}(ii), which shows the run of systematic (open square), statistical (pluses), and total uncertainty (filled circles) with \sigmasmooth. 

We select the optimal \sigmasmooth as that value that produces the smallest total uncertainty (the quadrature sum of the statistical and systematic uncertainties) and, thereby, provides the most reliable measurement of the TRGB discontinuity.
The distribution of values for our chosen \sigmasmooth is given in \autoref{fig:aslf_results}(iii) and the statistical and systematic uncertainties are $\sigma_{stat}$ = \trgbobsvalstaterrROUNDED~mag and $\sigma_{sys}$ \trgbobsvalsyserrROUNDED~mag, respectively. 
We note that these uncertainties are similar in magnitude to those originating from the calibration of the photometry described in \autoref{sec:phot}. 

\begin{figure} 
\includegraphics[width=\columnwidth]{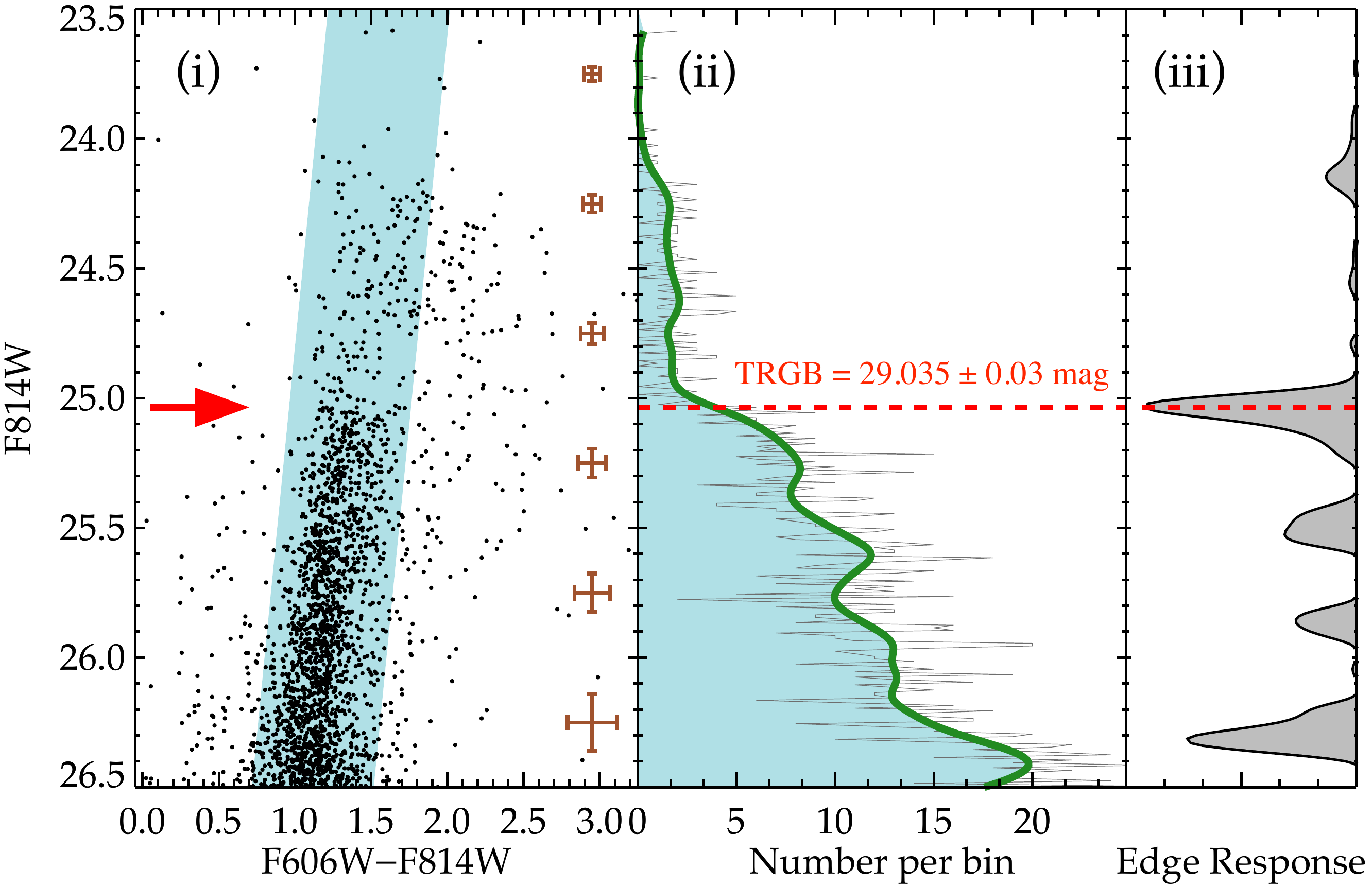}
\caption{\label{fig:our_trgb}
Detecting the TRGB discontinuity in \gal. 
(i) CMD of the CCHP field (M101\_B) with the color selection box in blue and the median magnitude and color uncertainties shown on the right side of the panel. 
(ii) The raw LF in bins of 0.01~mag is shown in gray with the GLOESS smoothed histogram (for \sigmasmooth=~0.05~mag) overplotted in green (and filled blue). 
(iii) The signal-to-noise weighted Sobel kernel response has a peak at \trgbobsvalROUNDED~mag, indicated in the red dashed line. 
The peak of the response function is also shown in (a) as the arrow and in (b) as the red dashed line.
We confidently detect the TRGB at \responsetoref{\trgbobsvalROUNDED~mag $\pm$ \trgbobsvalstaterrROUNDED~mag.}}
\end{figure} 

\subsection{Measurement of the TRGB Distance} 
Having determined the optimal smoothing scale and its measurement uncertainties for our field, we now can measure the apparent magnitude of the TRGB. 
This process is shown in the panels of \autoref{fig:our_trgb}.
Our color magnitude diagram (CMD) is shown in \autoref{fig:our_trgb}i with the median F814W magnitude and F606W-F814W color uncertainties shown on the right side of the panel.
Following \citetalias{hatt_2017}, the raw LF, binned at 0.01~mag, is shown in gray in \autoref{fig:our_trgb}ii with the GLOESS smoothed result overplotted (dark green). 
In \autoref{fig:our_trgb}iii, we apply the Sobel edge-detection algorithm and find the maximal response at, which is also indicated in Figures \ref{fig:our_trgb}i and \ref{fig:our_trgb}ii.
Based on this analysis, we detect the TRGB at \responsetoref{ \trgbobsvalROUNDED~mag $\pm$ \trgbobsvalstaterrROUNDED~mag in the F814W filter.}

\subsubsection{Galactic Extinction} 
We determine the reddening due to the Milky Way foreground using the online IRSA Galactic Dust Reddening and Extinction tool\footnote{\url{http://irsa.ipac.caltech.edu/applications/DUST/}} that queries the underlying \citet{schl98} maps and provides the \citet{sch11} rescaling for a location and radius of interest. 
We use the central coordinate of the CCHP field and use the average reddening, finding $E(B-V)$~= 0.0086~mag over a region 5\arcmin~in diameter \citep{sch11}.
Converting into the ACS filter system via \citet{cardelli_1989}, we find \Vextinction~mag for F606W and \Iextinction~mag for F814W, or $E(\mathrm{F606W}-\mathrm{F814W})$ = 0.008~mag.
Applying the foreground estimates to our TRGB measurement from the previous subsection, we find a foreground extinction-corrected TRGB magnitude of F814W~=~\trgbredcorrval~mag. 
Because the uncertainty in the color excess \citep[$\sigma_{E_{B-V}}\approx0.03$ via][]{schl98} is comparable to the reddening, we adopt half of the reddening as an additional systematic uncertainty.

We note that the reddening for M101\_B is marginally lower than the value at the center of \gal, where $E(B-V)$ = 0.018~mag \citep{sch11}.
Using NED, the central extinctions are A$_{F606W}$~=~0.051~mag and A$_{F814W}$~=~0.031~mag, which implies $E(\mathrm{F606W}-\mathrm{F814W})$~=~0.020~mag.
These differences are, however, within the $E(B-V)$ uncertainties of \citet{schl98} and within our estimate of the uncertainty at the position of our field.

\subsubsection{Internal Extinction}
Reddening internal to our \gal field is unknown, although there is some evidence for internal reddening in halos \citep[e.g., starting with][]{zaritsky_1994}.
More recently, \citet{peek_2015} used galaxies located behind stellar halos as ``standard crayons'' and was able to make a statistical assessment of the mean color-excess in stellar halos for galaxies at z$\sim$0.05 as a function of projected radius.
The \citeauthor{peek_2015} analysis resulted in a reddening profile that varied from 10 to 0.05 milli-magnitudes in $g-r$ over projected distances of 30~kpc to 1~Mpc.
The CCHP Field, however, is just internal to this profile, but the \citeauthor{peek_2015} analysis places a constraint on the likely reddening at the level of $\sim$0.01~mag.

On the other hand, the technique employed by \citeauthor{peek_2015} is statistical --- averaging the colors of many background galaxies and the specific properties of many host galaxies --- and, thus, it is not immediately evident how to use this result in application to a specific field in a specific galaxy.
In particular, there is not a good sense for the statistical distribution of dust in the halo, which could take two forms: (i) the filling factor of the dust in a full two-dimensional sense (e.g., the specifics of our pointing) and (ii) the variance of the reddening as a function of host-galaxy properties (e.g., the specifics of \gal). 

\responsetoref{In \citetalias{jang_2018}, \citetalias{hatt_2018a}, and  \citetalias{hatt_2018b}, we tested for the impact of halo reddening by} dividing the \hst+ACS pointing into subsections, detecting the TRGB in each, and evaluating differences in the measured TRGB magnitude, which could, in principle, be interpreted as being due to extinction differences across the field.
Unfortunately, we do not have sufficient source density in the M101\_A pointing to perform this test with rigor (i.e., the results are consistent within the noise). 
Based on the narrow RGB sequence in comparison to the color-uncertainties (\autoref{fig:our_trgb}), we suspect any reddening gradient to be small.

Thus, we have a potential systematic in our measurement, but one that we can anticipate to be small based on our data. 
We adopt an additional systematic uncertainty of 0.01~mag, based on the analysis of \citet{peek_2015} for their innermost bin, for the contribution of internal extinction in \gal.

\subsubsection{TRGB Absolute Magnitude}
\responsetoref{From Paper II to Paper V in the CCHP seriex, a provisional value  of \oldtrgblumwerr~mag determined in the Large Magellanic Cloud was adopted (note a slight adjustment was made in \citetalias{hatt_2018a} to better reflect the extinction as estimated in \citet[][]{hoyt_2018}).
In \citetalias[\freedmansub][]{Freedman2019}, a full re-analysis of the zero-point is presented, which includes a term for the $I$ to $F814W$ filter transformation. The zero point is \trgblumwerr~mag \citetalias[\freedmansub][]{Freedman2019}.
}
This TRGB absolute magnitude is broadly consistent with other TRGB calibrations \citep[e.g.,][among others]{rizzi_2007,janglee_2017}, with that determined from Galactic globular clusters ($M_{I}^\mathrm{TRGB}\approx-4$~mag), \responsetoref{and with the CCHP provisional value used in previous papers in this series.} 
The ultimate goal of the CCHP is to use \gaia trigonometric parallaxes to set a direct calibration of the absolute luminosity of the TRGB following the plan outlined in \citetalias{beaton_2016}.

\subsubsection{CCHP Distance to \gal}
Combining all of the terms determined in the previous subsections, we find a distance modulus to \gal of \truetrgbdmodwerr, or a distance of \truetrgbdmodMpcwerr. 
\autoref{tab_distance} summarizes the components of the distance and the associated uncertainties. 

\begin{deluxetable}{lccc} 
\tabletypesize{\normalsize}
\setlength{\tabcolsep}{0.05in}
\tablecaption{TRGB Distance and Error Budget to \gal \label{tab_distance}}
\tablewidth{0pt}
\tablehead{ \colhead{Parameter}  &  \colhead{Value} & \colhead{$\sigma_{ran}$} & \colhead{$\sigma_{sys}$} }
\startdata
TRGB F814W magnitude 	          & \trgbobsvalROUNDED  & \trgbobsvalstaterrROUNDED & \trgbobsvalsyserrROUNDED \\
$A_{\mathrm{F814W}}$	          & \IextinctioROUNDED        & \nodata            & \IextinctionerrTOTAL\tablenotemark{a} \\
\responsetoref{$M_{F814W}^{\mathrm{TRGB}}$} & \responsetoref{\trgblum \tablenotemark{b}}            & \responsetoref{\trgblumstaterr \tablenotemark{b}}    & \responsetoref{\trgblumsyserr \tablenotemark{b}} \\
\hline
True distance modulus [mag]       & \truetrgbdmod & \dmodcombinedstaterr & \dmodcombinedsyserr\\
{\bf Distance} [Mpc]              & \truetrgbdmodMpc & \truetrgbdmodMpcstaterr & \truetrgbdmodMpcsyserr\\
\hline \hline
\enddata
\tablenotetext{a}{Taken to be half of $A_{\mathrm{F814W}}$ and including a 0.01~mag component from internal extinction.}
\tablenotetext{b}{\responsetoref{\citet[\freedmansub][]{Freedman2019}}}
\end{deluxetable} 

\begin{figure} 
\includegraphics[width=0.7\columnwidth]{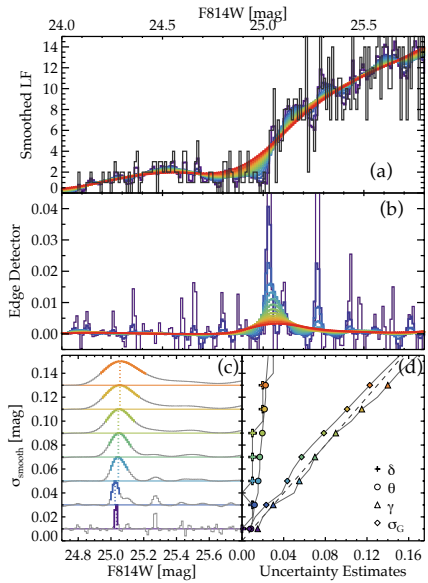}
\centering
\caption{\label{fig:our_trgb_stable} 
We study the reliability of our TRGB measurement as a function of the smoothing scale \sigmasmooth applied to the luminosity function. 
(a) The raw LF at 0.01~mag bins (black) is smoothed with successively larger \sigmasmooth, with blue being the smallest (0.01~mag) and red being the largest (0.14~mag).
(b) The edge-detection response ($\eta$) for each of the smoothed LF represented using the same color scheme as (a).
(c) A zoom into the response peak ($\eta_{max}$) for a subset of \sigmasmooth where $\eta$ has been normalized (grey). The full-width at half-max is shown thicker and color-coded as in (a). The magnitude of $\eta_{max}$, $m_{\eta_{max}}$, is indicated with a dashed vertical line for each $\eta$. 
(d) A comparison of different methods for estimating the precision of $m_{\eta_{max}}$ as a function of \sigmasmooth. 
Each symbols is a different quantitative assessment that are defined in the text:
 plus -- offset from the CCHP measurement ($\delta$),
 circle -- a metric similar to the error-on-the-mean ($\theta$) defined in the text by \autoref{eq:theta}, 
 triangle -- half-width at half-maximum ($\gamma$), and 
 diamonds -- dispersion for a Gaussian fit to the response function ($\sigma_{G}$), .
Taken together, the panels demonstrate that $m_{\eta_{max}}$ is determined to high precision (0.02~mag) irrespective of \sigmasmooth and that width of the response peak is not necessarily a good measure of the uncertainty on this determination.}
\end{figure} 

\subsection{Stability of the Edge-Detection}

The total number of stars at our TRGB detection is relatively small compared to previous papers in this series and it is not unreasonable to question if our chosen smoothing scale could be influencing our result.
To test for the impact of smoothing, we smooth the the raw luminosity function with a range of smoothing factors from \sigmasmooth=0.02~mag to \sigmasmooth=0.14~mag, apply our edge-detection kernel, and determine the TRGB magnitude from the peak of the response function. 
In several previous works the response function has been labeled ``$\eta$'', with the peak of the function being $\eta_{max}$ and the magnitude of the peak being $m_{\eta_{max}}$ and for clarity we will use this nomenclature in the discussions to follow \citep[e.g., starting with][]{lee_1993,madore_1995}.
The panels of \autoref{fig:our_trgb_stable} visually summarizes the result of this process, with \autoref{fig:our_trgb_stable}a being the progressively smoothed LF, \autoref{fig:our_trgb_stable}b showing $\eta$, \autoref{fig:our_trgb_stable}c a zoom into $\eta$ near $m_{\eta_{max}}$, and  \autoref{fig:our_trgb_stable}d exploring quantitative measures of the precision of $m_{\eta_{max}}$. 
Each of these panels will now be discussed in detail. 

In \autoref{fig:our_trgb_stable}a, the raw LF is shown in black, with the smoothed LF shown progressing from purple (\sigmasmooth=0.01~mag) to red (\sigmasmooth=0.14~mag). 
\autoref{fig:our_trgb_stable}a vividly demonstrates how the GLOESS algorithm progressively dampens the Poisson noise while also maintaining the broad features of the LF.
We can anticipate that $\eta$ will also show fewer noise spikes with increasing \sigmasmooth. 

Indeed, \autoref{fig:our_trgb_stable}b shows $\eta$ for the range of \sigmasmooth following the same color-coding as in \autoref{fig:our_trgb_stable}a and demonstrates how the numerous noise spikes are suppressed with \sigmasmooth. 
This suppression occurs {\it without loosing resolution in the LF} as would occur from using larger bin-widths. 
From inspection of \autoref{fig:our_trgb_stable}b, however, we can see that the dominant peak, at $m_{\eta_{max}}$ = \trgbobsvalROUNDED~mag, stays at nearly the same $m_{\eta_{max}}$ for all of the values of \sigmasmooth, but the value of $\eta_{max}$ itself decreases and the full-width of the peak becomes much broader with increasing \sigmasmooth. 
However, the $\eta$ for all \sigmasmooth, largely, has single dominant peak that is relatively isolated and determination of $m_{\eta_{max}}$ can be considered unambiguous for all values of \sigmasmooth. 

In \autoref{fig:our_trgb_stable}c, we zoom in on the x-axis to the $\sim$1~mag range around $m_{\eta_{max}}$ and the normalized $\eta$ for a subset of the \sigmasmooth values (e.g., such that $\eta_{max}$ = 0.02~mag on the y-axis for all \sigmasmooth). 
For each \sigmasmooth, $\eta$ is plotted in grey and $\eta_{max}$ is indicated with a vertical dotted line. 
From visual inspection, $m_{\eta_{max}}$ changes very little over the 0.14~mag range of \sigmasmooth.
The thicker colored portion of each response function illustrates the half-width at half maximum of the response peak ($\gamma$, hereafter), which becomes progressively broader with \sigmasmooth. 
In summary, we find little to no change in the $m_{\eta_{max}}$ with \sigmasmooth, although the value of $\eta_{max}$ decreases and the width of the response peak ($\gamma$) broadens with \sigmasmooth.
We quantify these observations in the following subsection. 

\subsection{Uncertainties on the Peak Magnitude} 
The panels of \autoref{fig:our_trgb_stable} are designed to study the impact of \sigmasmooth on the TRGB measurement for the CCHP field. 
Here we define and explore several quantitative metrics that attempt to measure the precision of $m_{\eta_{max}}$ from $\eta$ itself, rather than from simulations. 

In \autoref{fig:our_trgb_stable}d, the half-width at half-maximum, $\gamma_{\sigma_{s}}$ hereafter (demonstrated visually in \autoref{fig:our_trgb_stable}c) is plotted against \sigmasmooth with the triangle symbols (lines connect the symbols to guide the eye). 
Comparing the triangle symbols against the dashed one-to-one line, it is apparent that $\gamma_{\sigma_{s}}$ is strongly correlated with \sigmasmooth. 
The dispersion of a Gaussian profile fit to the $\eta$ near $\eta_{max}$ ($\sigma_{G,\sigma_{s}}$ hereafter) is shown by the diamond symbols in \autoref{fig:our_trgb_stable}d. 
Unsurprisingly, $\sigma_{G,\sigma_{s}}$ is also strongly correlated with \sigmasmooth.
Both $\gamma_{\sigma_{s}}$ and $\sigma_{G,\sigma_{s}}$ have been used in the literature to characterize the uncertainty $m_{\eta_{max}}$ directly from the edge response function, $\eta$ (and in consequence the precision of the TRGB determination).

In \autoref{fig:our_trgb_stable}d, the plus symbols show the absolute difference between the TRGB at the optimized smoothing scale of $m_{\eta_{max}}$ and that measured for a given \sigmasmooth ($\delta_{\sigma_{s}}$, hereafter). 
Mathematically, the term is 
\begin{equation} \label{eq:delta}
\delta_{\sigma_{s}} = \abs{m_{CCHP}-m_{\eta_{max}}(\sigma_{s})},
\end{equation}
\noindent where $m_{TRGB}$ is the CCHP TRGB magnitude ($m_{TRGB}=m_{\eta_{max}}(0.05)$=~\trgbobsvalROUNDED~mag) and $m_{\sigma}(\sigma_{s})$ is the peak of the response function for \sigmasmooth
The maximum value of $\delta_{\sigma_{s}}$ is 0.02~mag and it shows no correlation with \sigmasmooth.

The circle symbols in \autoref{fig:our_trgb_stable}d are a term ($\theta$ hereafter) that mimics the computation of the error on the mean.
More specifically, it is the full-width at half-maximum divided by the square root of the total number of counts contributing to the peak in the LF. 
Mathematically, $\theta$ is defined as follows:
\begin{equation} \label{eq:theta}
\theta_{\sigma_{s}} = \frac{\gamma_{\sigma_{s}}}{\sqrt{\sum_{\gamma_{-}}^{\gamma_{+}} (N_{*,i}) }}, 
\end{equation}
\noindent where $\gamma_{\sigma_{s}}$ is the half-width at half-maximum of the response peak, and $N_{*,m}$ is the value of the LF at bin $i$, and $\gamma_{-}$ and $\gamma_{+}$ define the magnitude bin for the half-width at half-maximum brighter and fainter than the peak, respectively. 
From inspection of \autoref{fig:our_trgb_stable}d, $\theta_{\sigma_{s}}$ is also uncorrelated from \sigmasmooth for the CCHP field.
Moreover, this term is much more similar to the values of the statistical uncertainty that we obtain from the ASLF simulations than either $\gamma_{\sigma_{s}}$ or $\sigma_{G,\sigma_{s}}$. 

We can compare these different means of assessing the uncertainty of the peak detection to the values determined from our ASLF procedure (\autoref{sec:aslf}). 
For the optimized \sigmasmooth, the statistical uncertainty is $\sigma_{stat}$ = \trgbobsvalstaterrROUNDED~mag and the systematic uncertainty is $\sigma_{sys}$ = \trgbobsvalsyserrROUNDED~mag. 
The $\delta$ term, which measures the difference of the response peak from peak in the optimized \sigmasmooth, is smaller than both our systematic and statistical uncertainty. 
Thus, for all \sigmasmooth we obtain a TRGB magnitude that is statistically consistent with the TRGB at the optimized \sigmasmooth. 
Thus, we conclude that the CCHP TRGB detection is insensitive to the smoothing applied to the LF.

In summary, we have explored the impact of GLOESS smoothing on the determination of the TRGB magnitude in the CCHP field. 
We measure statistically consistent TRGB values for \sigmasmooth between 0.01~mag and 0.14~mag and, thus, our measurement is not dependent on the smoothing scale.
For the LF in the CCHP field, we find that the width of $\eta$ near $m_{\eta_{max}}$ is entirely determined by \sigmasmooth and, as a result, this metric is not by design indicative of the precision of the TRGB determination from $\eta$ (e.g., we determine the same $m_{\eta_{max}}$ to within 0.02~mags for all \sigmasmooth). 
The caveat to this statement being that the CCHP field has an isolated RGB locus in the CMD that translates to single dominant response peak in $\eta$.
In the comparative analysis to follow in \autoref{sec:comptrgb}, we will demonstrate that this is not always the case.

\section{Comparative TRGB Analysis} \label{sec:comptrgb}

\gal is unique among the \sne-host galaxies in the CCHP in that, owing to its proximity to the Milky Way, there is ample archival data from \hst+ACS of comparable depth as our specially-designed pointing. 
Thus, in our study of \gal, we have the unique opportunity to test our measurement directly on other fields and to evaluate our field selection strategy in a quantitative fashion. 

\begin{figure*} 
\centering
\includegraphics[width=0.28\textwidth]{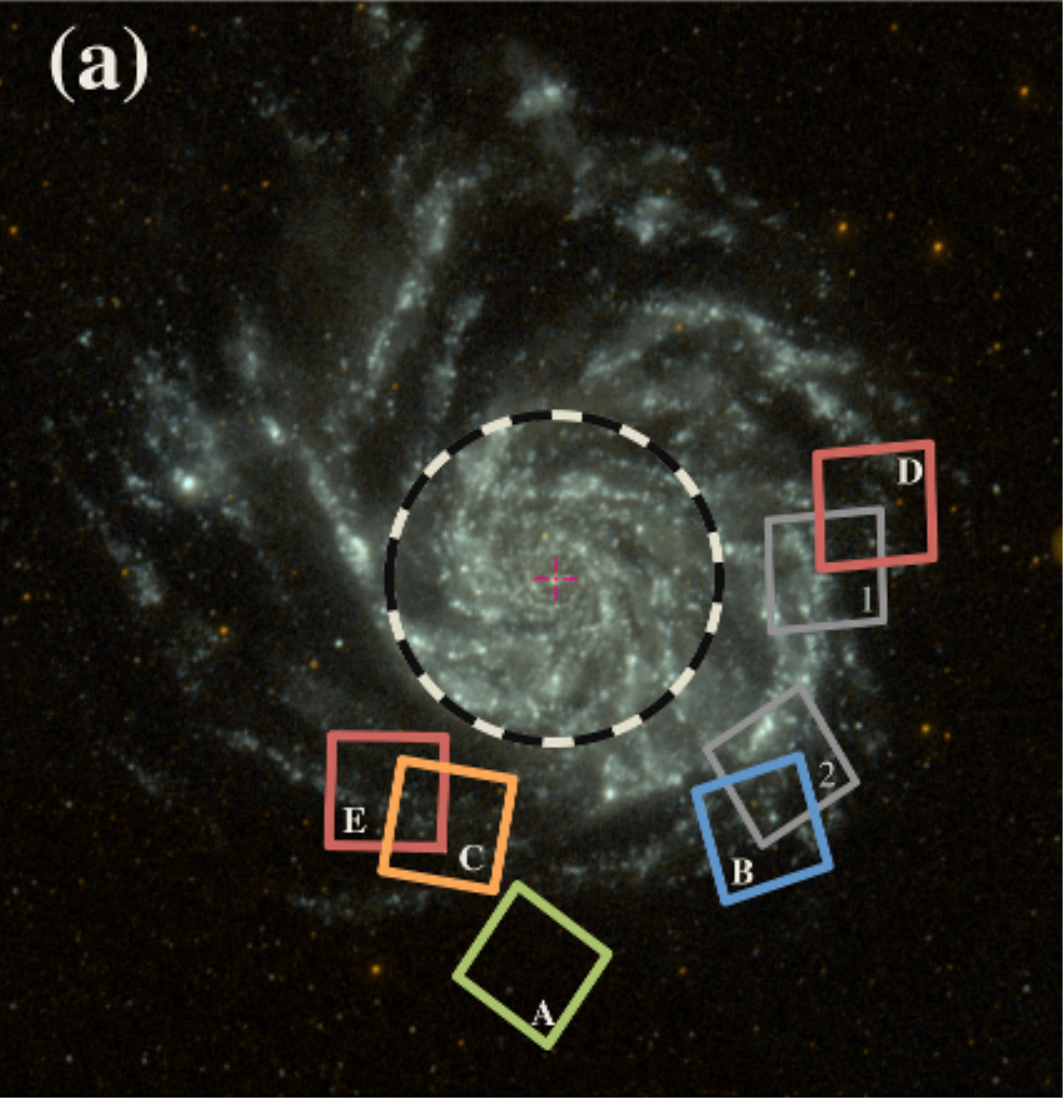}
\includegraphics[width=0.38\textwidth]{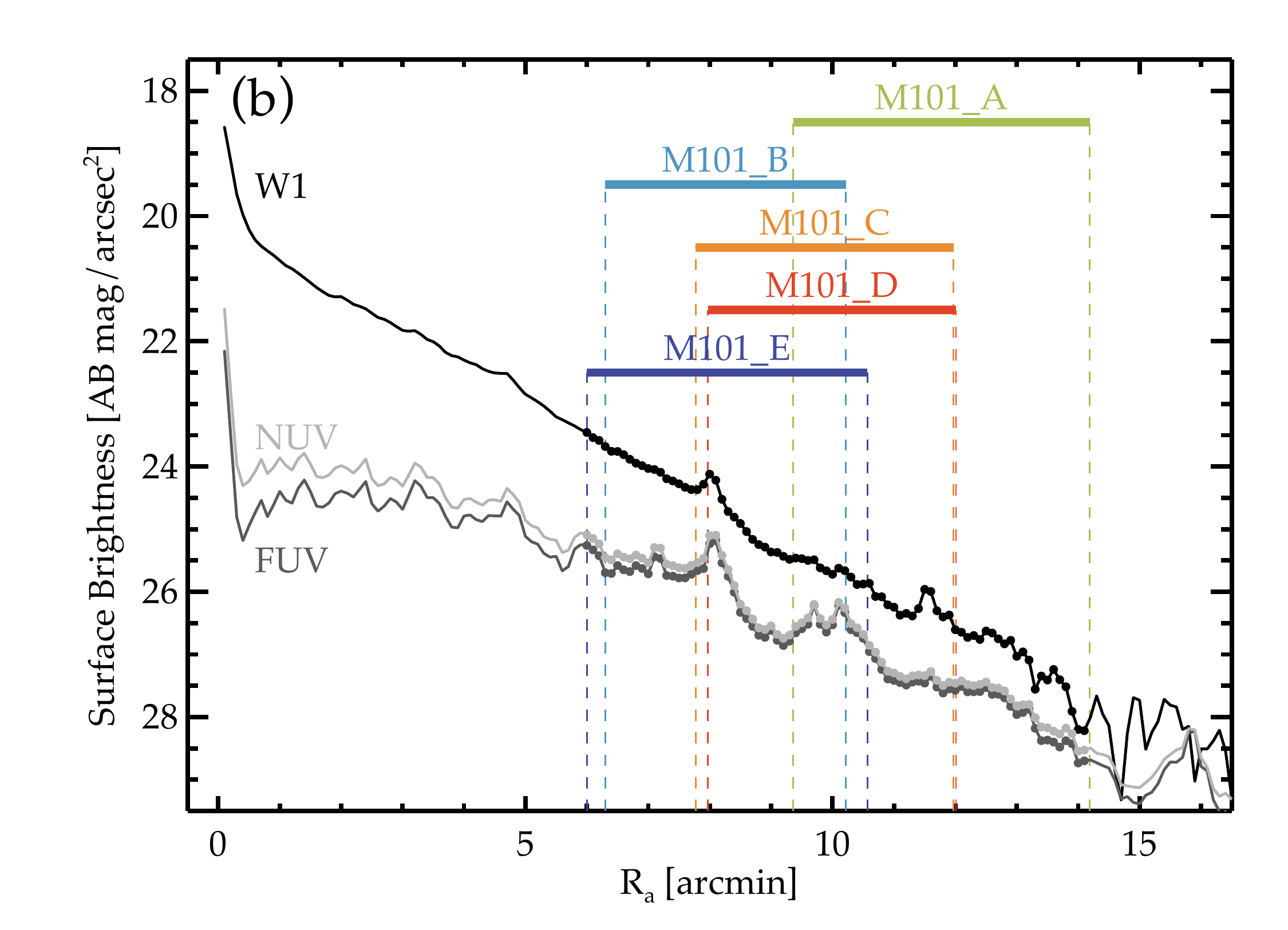}
\includegraphics[width=0.30\textwidth]{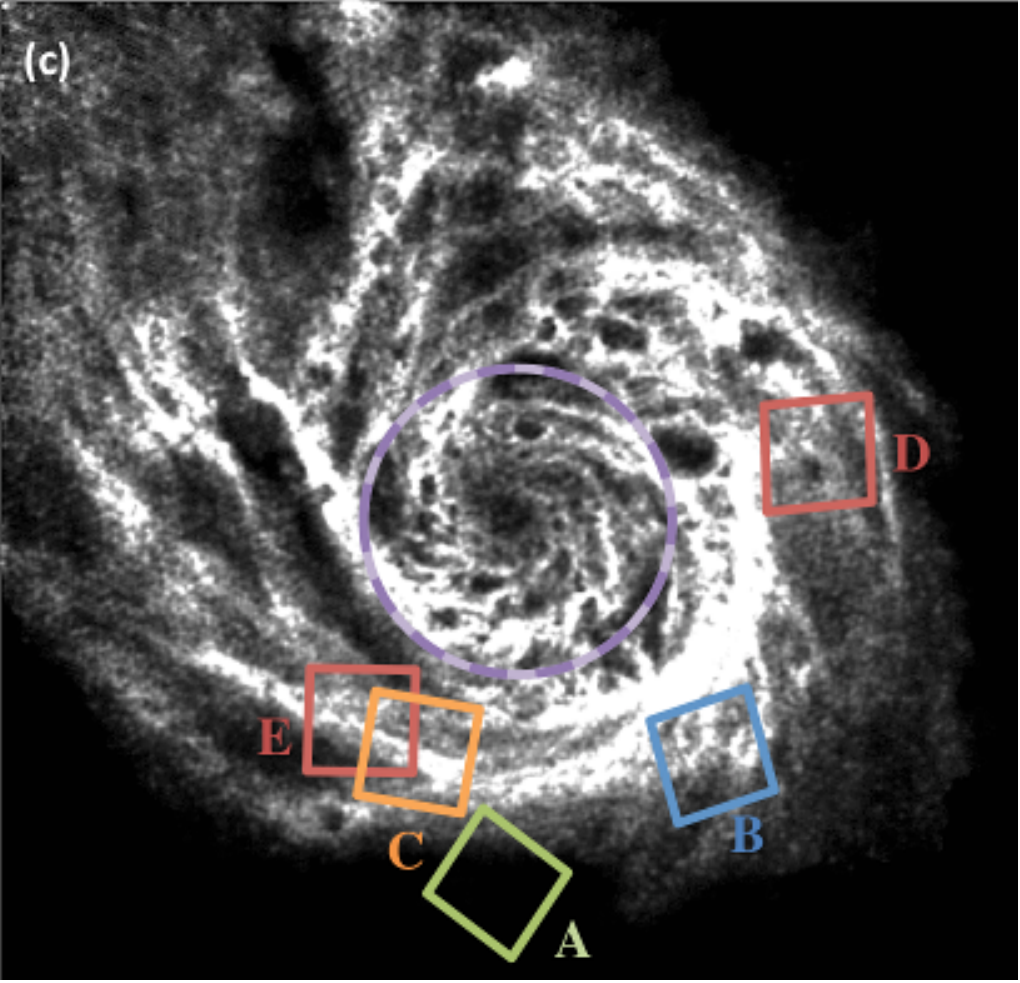}
\caption{ \label{fig:pointingmap_all} Location of the \hst+ACS imaging for \gal.
{\bf (a)}  Map of \hst ACS/WFC pointings in \gal that meet our signal-to-noise and crowding criteria for TRGB measurement (see \autoref{fig:pointingmap} for additional details).
The four archival pointings (labeled B through E and color-coded by the HST observing program) have been analyzed identically to the CCHP Pointing.
The two pointings indicated in grey (pointing 1 and 2) met our signal-to-noise requirements, but were not used due a high fraction of young stars. 
Technical details regarding the pointings are given in \autoref{tab:datasets}.
\textbf{(b)} Surface brightness profiles for M\,101 derived from \emph{GALEX} (NUV and FUV) and \emph{WISE} (W1) imaging. 
The radial extent for the five pointings used in our comparative analysis are shown.
The CCHP pointing (M101\_A) was chosen to span the 25-26~mag~arcsec$^{-2}$ isophote in W1 and the 27-28~mag~arcsec$^{-2}$ isophote in NUV. 
The CMD for the CCHP pointing (M101\_A; \autoref{fig:our_trgb}) suggests that these quantitative criteria select for Pop~II dominated regions. 
\textbf{(c)} \gal pointings used in this work overlaid on a neutral hydrogen map from THINGS \citep{walter_2008} and retrieved from NED. 
The overall scale is similar to that of \autoref{fig:pointingmap}a and \autoref{fig:pointingmap_all}a with the annulus representing a radius of 5\arcmin. 
Only the CCHP Field is outside of the gaseous disk of \gal (to the limiting column density of the observations).}
\end{figure*}

We searched the \hst archive for pointings with comparable depth and positioning, at least, in the outer disk of \gal. 
We restricted our search to ACS/WFC pointings, for which our pipeline has been optimized, and we also required F814W observations with an exposure time to attain S/N$\sim$10 at the approximate magnitude of the TRGB, and limited our search to pointings with central coordinates at R$_{\gal}$ \textgreater~ 5$\arcmin$ and having no dwarf satellite in the frame. 
Applying these conditions to the available imaging in the \hst archive, there were six suitable image datasets (not including the CCHP Field).
All six additional pointings are shown with the CCHP Field in \autoref{fig:pointingmap_all}a; these data are from GO14166 (Pointings C and 2), GO13737 (Pointing D), and GO13364 (Pointings E, F, and 1).
The color coding in \autoref{fig:pointingmap_all}a is coordinated with the \hst observing program.
Pointings 1 and 2 (indicated in gray in \autoref{fig:pointingmap}a) are both visually dominated by spiral arms and were deemed not suitable for our goals.
The remaining four were considered suitable, though not ideal, for a TRGB measurement and their observation details are given in \autoref{tab:datasets}.

First, we compare the locations of these fields to 1-D and 2-D maps of \gal in \autoref{sec:arc_sb}. 
Then we compare the CMDs of the archival fields to that of the CCHP Field in \autoref{sec:arc_cmds}. 
A detailed TRGB detection analysis is given in \autoref{sec:arc_trgb} and a discussion of the internal extinction is given in \autoref{sec:arc_ext}. 

\begin{figure*} 
\includegraphics[width=0.6\textwidth]{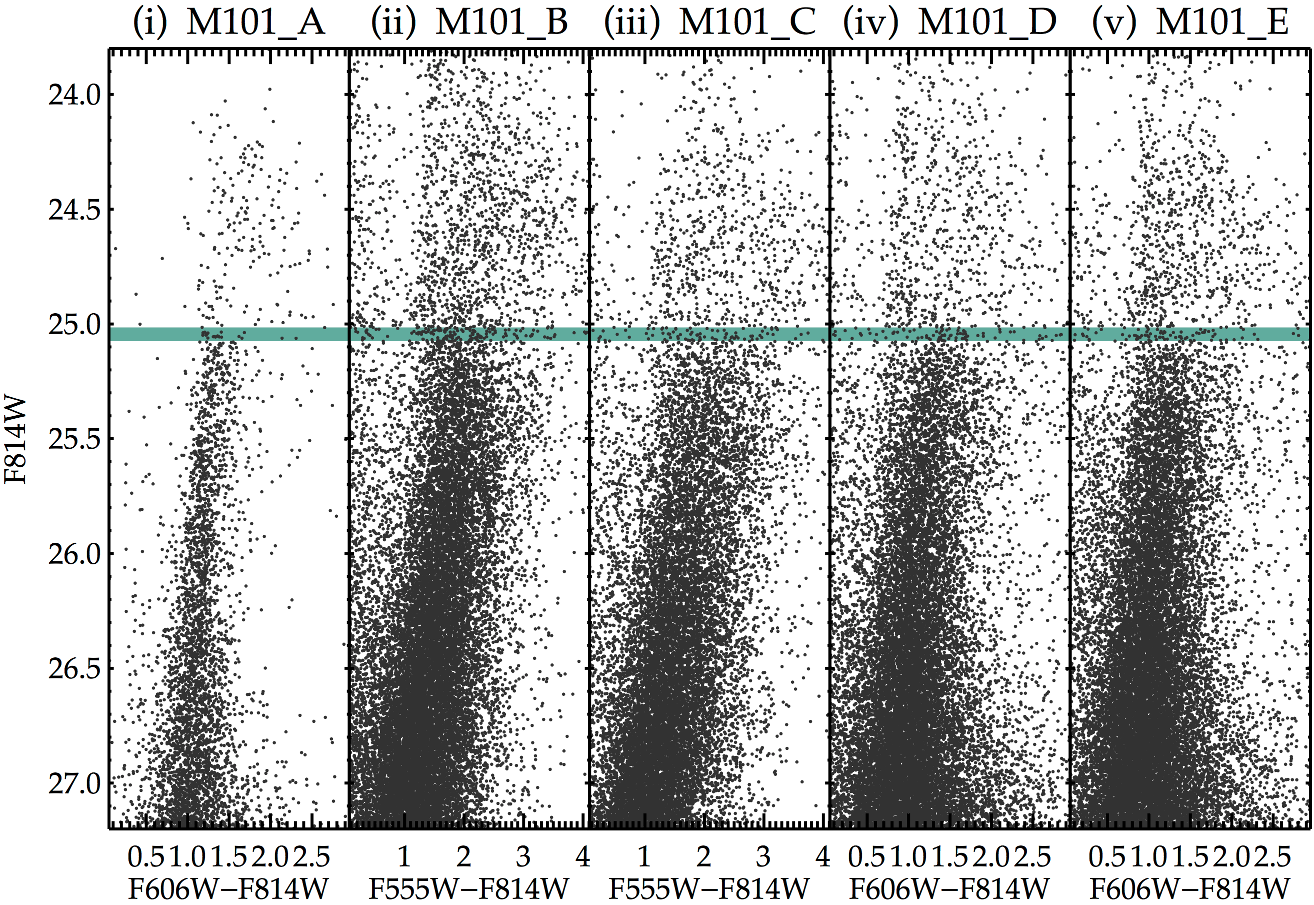}
\caption{ \label{fig:cmds}
Color-magnitude diagrams for the five fields in the outer-disk of M\,101 used in this work. 
The panels are 
 (i) our `pure halo' pointing for the CCHP (M101\_A), (ii) the M101\_B pointing, 
 (iii) the M101\_C pointing, (iv) the M101\_D pointing, and (v) the M101\_E pointing.
Note that the color in panels (i), (iv), and (v) is $F606W$-$F814W$
 and in panels (ii) and (iii) is $F555W$-$F814W$, with the x-axis color range spanned by the former being transformed into the latter system as described by \autoref{eq:colortransform}. 
A filled band represents the TRGB discontinuity determined from M101\_A with the width of $\pm$1-sigma (Table~\ref{tab_distance}).  
The difference in the stellar populations amongst the fields is visually striking and largely conforms to the placement of the pointings in relation to the spiral arms as demonstrated by \autoref{fig:pointingmap_all}a.}
\end{figure*} 

\subsection{Comparison to Surface Brightness Profiles} \label{sec:arc_sb}
Across the CCHP, the \hst+ACS imaging fields were selected based on a set of quantitative surface brightness criteria using a combination of archival \wise and \galex imaging. 
More specifically, each pointing was selected to straddle the 27-28~mag~arcsec$^{-2}$ isophote while also being on the 25-26~mag~arcsec$^{-2}$ isophote in the \emph{WISE}-1 band (W1). 
The combination of these surface brightness criteria was meant to mitigate contamination from younger populations (traced by the UV) while also having enough RGB stars to to able to make our measurement (traced by W1). 
Where necessary, this overall strategy was modulated to avoid extended disks (especially important for \gal) and other young or tidal structures obvious in archival imaging; as a result, the fields were often positioned on the minor axis of the galaxy. 

In \autoref{fig:pointingmap_all}b, we compare the approximate locations of the five \hst+ACS fields to \wise-W1 and the \galex-FUV and -NUV surface brightness profiles. 
These surface brightness profiles were constructed similar in form to the methodology presented in the \galex Nearby Galaxy Atlas \citep{gildepaz_2007}, but were applied to custom builds of the \galex and \wise imaging for \gal. 
The approximate radial extent for each of the five pointings is indicated with the colored bars, with the CCHP Pointing (M101\_A) spanning both the largest radial extent and being the most distant from the \gal center.
Each of the archival pointings fall on visible features in the \galex surface brightness profile that indicate spiral arms and, by association, contaminating young and intermediate-aged populations that could systematically bias an otherwise precise detection of the TRGB.
Thus, we would predict from \autoref{fig:pointingmap_all}a and \autoref{fig:pointingmap_all}b that the archival pointings, despite having strong RGB sequences as indicated by a higher \wise W1 surface brightness, will suffer from disproportionately stronger contamination by young and intermediate age populations.
This contamination has the potential to make the accurate and precise measurement of the TRGB difficult, despite there being more RGB stars.

\subsection{The Color-Magnitude Diagrams} \label{sec:arc_cmds}

The images for the archival fields were processed and photometry was produced identically to that for the CCHP field (\autoref{sec:phot}), but with appropriate modifications of the photometric zero points (these are given in \autoref{app} in \autoref{tab:calibration}). 
We restrict the photometry for each pointing using the scheme described for our CCHP field (\autoref{fig:phot_quality}), 
but with an adjustment made, based on the photometric depth of the field, to the magnitude at which the photometric error model transitions from a constant to an exponential form (visualizations can be found in \autoref{app} in \autoref{fig:phot_qual_all} and values are given in \autoref{tab:calibration}).

\autoref{fig:cmds} presents the color-magnitude diagrams for each of the five fields. 
To compare the F555W-F814W color scales (pointings M101\_B and M101\_C), we transform the F606W-F814W color ranges to the approximate F555W-F814W color following the relationships defined by \citet{janglee_2017} as follows:
 \begin{multline} \label{eq:colortransform} 
  \mathrm{F555W}-\mathrm{F814W}= \\
  1.393 (\pm 0.003) \times (\mathrm{F606W}-\mathrm{F814W}) - 0.004(\pm 0.004) 
 \end{multline} 
for stellar sources with (F606W-F814W) \textless~1.5 mag.
Using the transformation, the x-axis scale for the panels of \autoref{fig:cmds} was modified for M101\_B and M101\_C such that any apparent differences on the RGB are not driven by the physical differences in the filter systems (F555W-F814W is naturally a larger color baseline than F606W-F814W).
The green band in \autoref{fig:cmds} gives the TRGB magnitude determined for the CCHP field, with the width indicating its statistical uncertainty. 

Comparing the CCHP Field (\autoref{fig:cmds}i) to the other four fields, there are a number of noteworthy differences.
First, the total number of stars is dramatically higher in the archival fields for all stellar types.
Second, there are pronounced young (blue) and intermediate-aged (AGB) populations in the archival fields that are absent (or nearly so) in the CCHP field.
Third, the RGB in the archival fields is much broader in color, which can be attributed to these fields sampling stellar populations with a broader range of ages and metallicities, as well as possibly being due to differential extinction within the structural components probed by these lines of sight. 

The impact that these differences have on the TRGB identification is not immediately obvious from the panels of \autoref{fig:cmds}. 
Visually, the discontinuities do not look significantly different from that in the CCHP field (green band), albeit these look slightly fainter. 
While the contaminating populations are a concern, it could be argued that the significantly larger number of RGB stars contributing to the TRGB would counterbalance the impact from the other populations.
In the next subsection, we explore quantitatively why this is not the case.

\begin{figure} 
\centering
    \includegraphics[width=1.0\columnwidth]{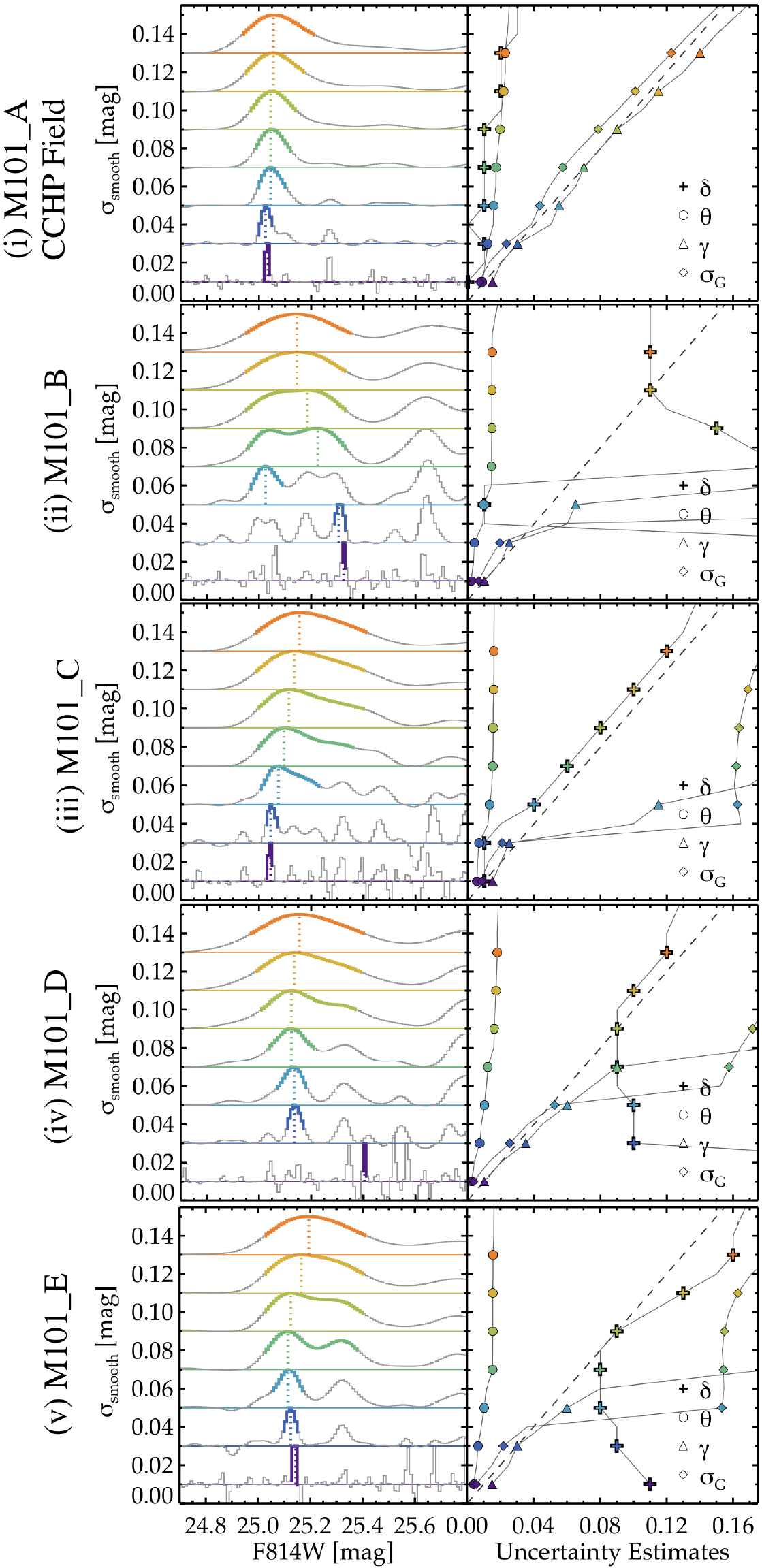}
\caption{\label{fig:smoothing} Edge-detections (left) and uncertainty estimates (right) as in \autoref{fig:our_trgb_stable}c and \autoref{fig:our_trgb_stable}d, respectively, for each of our five fields, from top (i) M101\_A, (ii) M101\_B, (iii) M101\_C, (iv) M101\_D, and (v) M101\_E. The detection stability seen for the CCHP Field is not see in the archival fields as their edge detection response functions are broader, multi-peaked, and asymmetric. Moreover, common metrics to define the ``precision'' of result via the edge response show little to no correspondence with the actual uncertainty.}
\end{figure} 

\subsection{TRGB Determination in Archival Fields} \label{sec:arc_trgb} 
In demonstrating the stability of our TRGB detection for the M101\_A field, we undertook an exercise of comparing the LF, edge-response, and predicted uncertainties for a range of smoothing factors.
We repeated this procedure for each of the four archival fields; 
\autoref{fig:smoothing} contains the summary panels for each of the fields (\autoref{fig:our_trgb_stable}c and \autoref{fig:our_trgb_stable}d) and the full visualizations are given in \autoref{app}, \autoref{fig:smoothing_full}. 
The colors, annotations, and axis ranges are identical between the panels and the M101\_A field is repeated for ease of comparison.

A comparison of the LF for each of the fields quantifies the considerably larger number of stars in the archival fields for all magnitudes. 
However, while the RGB population improved by roughly a factor of 3, the contaminating AGB population increases in number by roughly a factor of 5 to 6. 
Nevertheless, from visual inspection the sharpest TRGB jump in the LF is in M101\_A, with the archival fields showing a much more gradual transition from the AGB to RGB. 
The GLOESS smoothing makes the transition even more gradual but, as we observed in application to the CCHP field, it does not innately distort the LF shape.

Comparing the edge response ($\eta$) for the archival fields to the CCHP field begins to reveal how the contaminating populations impact the determination of the TRGB (\autoref{fig:smoothing}, left panels).
At the smallest smoothing scales (less than 0.02 mag, dark blue/purple), the edge detection response has significantly more false-peaks occurring both brighter and fainter than the TRGB; these discontinuities are due to Poisson fluctuations in the LF.
Sometimes, multiple peaks occur very near each other making the selection of the most appropriate peak difficult.
On the other hand, while larger smoothing scales serve to dampen these peaks, the peaks also start to merge with each other producing composite peaks in the edge response ($\eta$) that are both non-Gaussian and asymmetric.  

The consequence of the latter point is demonstrated quite clearly when comparing $m_{\eta_{max}}$ (vertical dotted lines) for each smoothing scale in \autoref{fig:smoothing} (left panels).
The effect is most striking for \sigmasmooth of 0.06~mag and larger for M101\_C and M101\_D (\autoref{fig:smoothing}(ii) and \autoref{fig:smoothing}(iii), respectively), where the width of the peaks is quite asymmetric and broad (spanning nearly the full range of the figure.).
Furthermore, the maximal response, itself, can shift as much as 0.3~mag across 0.01~mag smoothing shifts.
In addition to sharp jumps in the maximal response with smoothing factor, we also see slow drift of the maximal response (\autoref{fig:smoothing}(iii) and \autoref{fig:smoothing}(v)) and this drift can be both to fainter or brighter magnitudes (e.g., \autoref{fig:smoothing}(ii)).

One could attribute the instability of the $\eta_{max}$ to either the method or size of the smoothing. 
In the CCHP field, however, we show that we have a stable response for all smoothing scales, including that of ``no smoothing'' (\autoref{fig:smoothing}i). 
Stated differently, for the CCHP field the smoothing serves to clarify the result and not to bias or distort it. 
Instead, we attribute the source of the instability in the archival fields to the contaminating populations that have the effect of adding considerable noise to the LF both brighter and fainter than the TRGB. 
Moreover, we are seeing composite populations along the line of sight; while their physical separation is not resolved at the precision of our photometry, the differences in the dust column between their physical separations could produce noise spikes from bona-fide TRGB stars. 
This noise, in turn, can both mask the true TRGB peak and, with larger smoothing scales, ``drag'' the peak away from the TRGB.  

Another aspect of the TRGB detection via the maximal response function is also worth mentioning. 
It is often argued in the literature that the ``closest'' TRGB stars will contribute to the peak and thus complexity along the line-of-sight is not a concern \responsetoref{\citep[see e.g.,][among others]{shappee_2011,Tikhonov_2015}.} 
As our experiment with the archival fields has shown, however, the maximal peaks most often correspond to where the ``bulk'' of the TRGB stars are along-the-line of sight into the galaxy. 
At small smoothing scales we do often get peaks at or near the ``halo'' pointing of M101\_A, but as smoothing is incrementally applied, the peaks tend to be at fainter magnitudes, which, as previously discussed, are likely TRGB stars in the disk of \gal,  are consistent with being behind some dust column, and likely comprise a more complicated mix of stellar populations than our halo sample (e.g., metallicity/age effects). 
We will return to this point in \autoref{sec:disc} with respect to literature studies of \gal applying the TRGB method.

\responsetoref{We also note how the uncertainty estimates in the right panels of \autoref{fig:smoothing} change with the indeterminate results in the right panels. 
The quantity $\delta$ defined in \autoref{eq:delta} as the difference between the CCHP measurement and that in an individual case behaves erratically as anticipated, with the exception of \autoref{fig:smoothing}iii and \autoref{fig:smoothing}iv, where at some point it grows in lockstep with the smoothing factor (it traces the 1:1 line in right the panels).
Of note, the quantity $\theta$ (circles) defined in \autoref{eq:theta} shows consistent behavior for all five panels, meaning that it is a poor measure of the distance uncertainty. 
Likewise, the quantities $\gamma$ and $\sigma_{G}$ seem to also perform poorly as measures of the uncertainty, as they either grossly over- or under- estimate $\delta$, but are always equal to or larger than the smoothing scale, \sigmasmooth.
At some level, however, they are representative of how reliable a heavily smoothed LF is, in that at large smoothing scales these estimates do provide 1-$\sigma$ consistency with the CCHP result -- this is seen in \autoref{fig:smoothing} as $\delta$ is smaller than $\gamma$ or $\sigma_{G}$ for \sigmasmooth~\textgreater$\sim$0.05 mag. 
We conclude here in most cases, error estimates that are derived from the response function $\eta$ are not reliable measures of the uncertainty with the exception of $\gamma$ and $\sigma_{G}$ in the cases of large smoothing (\sigmasmooth~\textgreater$\sim$0.05 mag). This does provide a cautionary note to error estimates that are significantly smaller than the effective smoothing scale of the LF, whether that smoothing is conducted by binning, smoothing of the LF, or broadening of the edge-detection kernel, in that the response $\eta$ is heavily impacted by these choices.}

Thus, while our original goal was to compare the measurements between archival fields and the CCHP field, we have, instead, come to the conclusion that we cannot unambiguously measure a distance to any of the archival fields at comparable certainty. 
Moreover, our ASLF methods that were designed to determine the ``optimal'' smoothing scale for measuring the TRGB and its associated uncertainties are not well suited to handle the complexity of the stellar populations in the archival fields.
Thus, for rigorous, high-precision detection of the TRGB (as is our goal here), we come to the conclusion that the selection of an appropriate field is as important, if not more important, than considerations for the number of stars and the depth of the imaging.

\subsection{Extinction in the Archival Fields} \label{sec:arc_ext} 

In the previous subsection, we used archival HST data to explore the impact of ambiguous peaks in the edge response function on the measurement of the TRGB. 
However we also need to consider differences in the total extinction along the line-of-sight.
The extinction from the Milky Way foreground is very similar between all five fields (the exact values for each field are given in Table \ref{tab:calibration} and were determined identically to that of the CCHP field), but we must also consider the potential for additional internal-extinction within \gal.
More specifically, the archival pointings look into the outer disk of \gal and thereby probe multiple structural components; in the Milky Way, a similar pointing would probe the halo (and potentially substructure) and both the thick and thin disk. 
An additional complication, however, will be that each of these structures will have their own associated columns of interstellar material, with the thin disk being the most likely to have appreciable amounts, while also having the largest number of stars. 
Taken together, we anticipate that we are most likely to trigger off of a thin disk population which would, by design, require an additional term for extinction within \gal. 

In comparing the TRGB identifications from the maximum response in \autoref{fig:smoothing}, the majority of the detections are just fainter than that in the CCHP field --- with the exception of the small smoothing scales (\sigmasmooth \textless 0.04~mag) for M101\_D in \autoref{fig:smoothing}(iii) and intermediate smoothing scales for M101\_C \autoref{fig:smoothing}(ii).
While the archival pointings are in the outer regions of \gal, they are still in regions with significant UV emission (\autoref{fig:pointingmap_all}a and \ref{fig:pointingmap_all}b) and show signs of relatively young stellar populations (\autoref{fig:cmds}). 

In \autoref{fig:pointingmap_all}c, the positions of the HST fields are overlaid on the total power HI map from the THINGS survey \citep[][retrieved from NED]{walter_2008}. 
The CCHP field (M101\_A) is largely free of gas to the limiting column of the survey, which is $\sim$3.2$\times$10$^{20}$ cm$^{-2}$ in the ``normal'' spatial resolution \citep{walter_2008}. 
The conversion from atomic gas to reddening is quite nuanced, especially so at the few 0.01~mag level, but that the projected locations of the archival fields are all within the gaseous disk provides circumstantial evidence of there being internal extinction for the RGB stars in the disk.
Because the highest density of RGB stars along the line-of-sight is in the disk for these archival pointings, some fraction of those RGB stars will be impacted by the presence any dust associated with the HI gas.
\responsetoref{Moreover, high-quality HI maps are often available for nearby galaxies and this is an aspect of field selection that can, and should, be considered for high-precision TRGB measurements. 
We note that this guideline can be applied both to select new or to screen archival observations; the latter being performed in \citet{anand_2018}.}

Thus, the observation that most of the TRGB measurements made in the archival fields are fainter than that in the CCHP field is fully consistent with there being extinction from interstellar material in outer disk of \gal.
Or at least that the dominant contribution to the TRGB signal is a population of stars behind some amount of dust. 
Indeed, some of the ambiguity discussed in the previous section could be coming from RGB stars in multiple structural components of \gal behind different amounts of dust. 
By moving \responsetoref{safely} into the stellar halo, our TRGB measurements are less susceptible to systematic effects from dust extinction. 

\subsection{Discussion} 

In this section, images for four archival points in \gal were processed and analyzed identically to that of the CCHP field.
We demonstrated that we cannot determine a TRGB detection in any of these fields unambiguously. 
Moreover, we have shown that our pointings simultaneously avoid contamination from younger stellar populations and minimize interstellar extinction effects to a level below our measurement uncertainties.
\responsetoref{We posit that the quantitative field selection strategy of the CCHP has helped its success in this regard.}

\responsetoref{Qualitatively, the CCHP targeted fields in the ``stellar halo.''  We note that other programs used a similar selection strategy by focusing on the outer-components of galaxies \citep[most notably,][among others]{McQuinn_2016a,McQuinn_2016b,McQuinn_2017}. Inspection of these works indicate, that although these fields are outside of the traditional disk, they contain contamination from stars more luminous than the TRGB -- both AGB stars or red super-giant sequences \citep[for example M\,74 in ][]{McQuinn_2017}. Moreover in application to \gal, we have shown that fields in its outer regions they still have an unknown, and indeed perhaps \textit{unquantifiable}, impact from interstellar extinction. The presence of young to intermediate age populations and not insignificant extinction were anticipated in the CCHP in part due to detailed observations of the outer Milky Way disk that show these characteristics to large radii \citep[for a brief overview see][]{carraro_2015}. }

\responsetoref{ Quantitatively, the CCHP performed targeting using wide-area, multi-wavelength surface-brightness maps from well characterized all-sky surveys. Of particular utility was placing a field that sampled a specific isophote in \galex (that track young populations even to low stellar density) and \wise (that track the older stellar populations) using two-dimensional maps; this is demonstrated by the radially binned map shown in \autoref{fig:pointingmap_all}b. These quantitative choices were found by a review of literature CMDs and their placement on surface-brightness maps \citep[a good example being the CMD of NGC\,4258 in][]{Mager_2008}. Likewise, HI maps \autoref{fig:pointingmap_all}c provided additional diagnostics to avoid interstellar material. The quantitative metrics demonstrated in \autoref{fig:pointingmap_all} were used for all galaxies in the CCHP sample \citepalias[see][]{beaton_2016}. }

\section{Discussion} \label{sec:disc}

\begin{figure} 
\centering
\includegraphics[width=0.8\columnwidth]{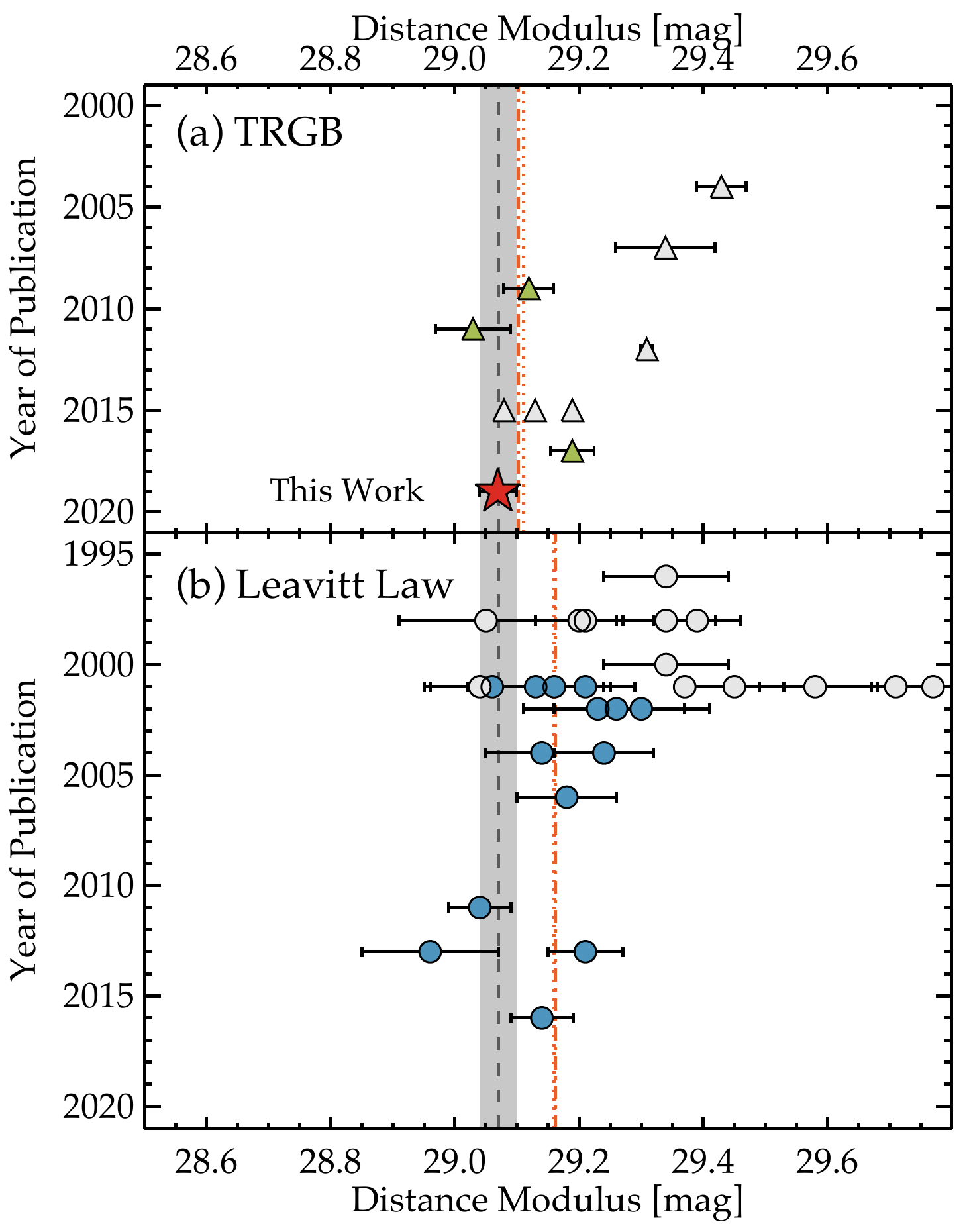}
\caption{\label{fig:distcomp}
Comparison of literature distances to \gal using (a) the TRGB and (b) the Leavitt Law for Cepheids. 
In both panels, the value for the CCHP and its statistical uncertainty are shown as the dashed line and gray band, respectively.
For the TRGB measurements, we have homogenized the absolute magnitude of the TRGB to the value used in our work. 
For the Leavitt Law distances, the values were compiled from NED \citep{steer_2017} and no homogenization has been applied owing to the numerous terms that need to be taken into account, although we have attempted to use those on an LMC-based scale for consistency with the TRGB. 
In each panel, the red dotted line and dot-dashed lines show the uncertainty weighted and unweighted means, respectively, based on the filled symbols (green for the TRGB and blue for the Leavitt Law). 
See the text for more discussion.
} 
\end{figure}

Our primary purpose in evaluating the archival fields in this work was to understand if or how our TRGB measurement compares to other locations in \gal.
We now compare these measurements to those from earlier studies.
Instead of comparing the final distance moduli, we will compare the actual tip magnitudes and the locations of the fields used in each of the papers to eliminate differences in the adopted TRGB zeropoints.
As demonstrated in previous discussions, the variation of the foreground Milky Way extinction across \gal is small and unlikely to be a large concern in the comparisons to follow, whereas extinction \textit{internal} to the component of \gal being studied is more problematic.
The distance moduli and uncertainties are visualized in \autoref{fig:distcomp}a using our TRGB zeropoint. 
The data used in \autoref{fig:distcomp} is given in \autoref{app:litdistances}, with references to all studies not mentioned explicitly in the discussion to follow. 

\subsection{Summaries of Previous TRGB Studies}

 \citet[][S04]{sakai_2004} used the two outer chips of a WFPC2 pointing in the outer disk of \gal  \responsetoref{following the data-analysis strategy of \citet[][from the Key Project]{Hill_1998}}.
The authors use a series of edge-detectors, including a linear LF, log LF, and a cross-correlation technique.
The authors find overall consistent tip detection at $I$=25.41~$\pm$~0.04 mag. 
They comment on the field being contaminated heavily by young- and intermediate-aged populations. 

\citet[][R07]{rizzi_2007} repeated the analysis of S04, but using the HSTPhot package \citep{dolphin_2000} to produce photometry of the same fields. 
The authors determined a TRGB magnitude of $I_{TRGB}$ = 25.31 $\pm$ 0.08 mag, which is in rough statistical agreement with that derived by \citetalias{sakai_2004} \responsetoref{(there is a $\sim$1$\sigma$ difference).}

\citet[][SS11]{shappee_2011} used two HST+ACS fields in the inner disk of M\,101 (within the 5$'$ radius plotted in \autoref{fig:pointingmap_all}a) \responsetoref{and determined photometry using the DOLPHOT package that is specifically optimized for crowded field photometry.}
Of note, DOLPHOT applies CTE corrections internally and derives its own aperture corrections.
\citetalias{shappee_2011} only use sources that are more than 4.75$'$ in radial separation from \gal (to reduce AGB contamination) and remove sources with $V-$\textless~1.0~mag (to reduce contamination both from blue AGB stars and younger populations). 
\citetalias{shappee_2011} adopt the ``continuous'' form of the LF from \citetalias{sakai_2004}, the logarithmic edge-detector from \citet{mendez_2002}, and apply a Poisson-statistics based signal-to-noise weighting scheme. 
At its spirit this TRGB detection formulation is the most similar of the literature studies to that undertaken in our CCHP Field (M101\_A).
Recognizing that their field will contain a range of metallicities, \citetalias{shappee_2011}  translate the photometry into the $T$ magnitude defined by \citet{Madore_2009}, a magnitude system which was designed to ``remove'' the downward slope of the TRGB for a multi-population RGB \citep[a comparison of this technique and others is given][their section 4.2.2 and figure 28]{beaton_2018}.  
\citetalias{shappee_2011} obtained a tip magnitude of $T$ = 25.00 $\pm$ 0.06 mag, which is notably different from the values derived by \citetalias{rizzi_2007} and \citetalias{sakai_2004}, above. 

\citet[][LJ12]{lee_2012} re-reduced eight fields, including those used by both \citetalias{sakai_2004} and \citetalias{shappee_2011}. The authors consistently found a TRGB magnitudes between 25.24 and 25.30~mag with typical uncertainties of order 0.03~mag. The $T$ magnitudes range from 25.15 to 25.33~mag with similar uncertainties. The values from \citetalias{lee_2012} are in agreement with those of \citetalias{rizzi_2007} and \citetalias{sakai_2004}, and differ by 3-$\sigma$ to 5.5-$\sigma$ from that of \citetalias{shappee_2011}. 
These are all pointings well within the extent of the UV and gaseous disk of \gal.

In \citet[][T15]{Tikhonov_2015}, three \hst+~ACS fields were analyzed, one of which is  M101\_D. 
The TRGB detection in each field was as follows: F1: 25.05 mag, F2: 25.10 mag (Pointing 1), and F3: 24.11 mag (M101\_D). \citetalias{Tikhonov_2015} provide no uncertainties and the details of the exact proceedures employed in this work are more sparse than those in other literature studies discussed here. 
Qualitatively, these results agree with our processing in these fields. 
The final distance quoted by \citetalias{Tikhonov_2015} appears to be that of their F3, our M101\_D, field, which is at the largest projected radial distance from \gal.

In \citet[][JL17]{janglee_2017}, the \gal measurements of \citetalias{lee_2012} were revisited by the authors. 
\citetalias{janglee_2017} use the pointing we call M101\_E (see \autoref{tab:datasets}) and find a TRGB magnitude of $\mathrm{F814W} = 25.16\pm0.035$~mag. 
This is 0.12~mag brighter than the result in LJ12. 
In their section 4.1, \citetalias{janglee_2017} explained the difference between this result and \citetalias{lee_2012} as being due to (i) changing the procedure for the aperture correction and (ii) having used images that were not fully corrected for the charge-transfer-efficiency. 
In addition, \citetalias{janglee_2017} note that the M101\_E data was significantly deeper than that used in \citetalias{lee_2012} and had a visibly stronger TRGB signal, both of which would impact the result. 
\citetalias{janglee_2017} binned their data in 0.05~mag bins and applied a ``classic'' Sobel kernel ([-1, -2, 0, +2, +1]) to find a broad edge-response peak between 25.0 and 25.3 mag with a similar TRGB detection in both the F814W and $QT$ photometric systems; the latter is a variant of the $T$ system that incorporates a quadratic color-term \citep[a comparison of this technique and others is given][their section 4.2.2 and figure 28]{beaton_2018}.

Lastly, the Extragalactic Distance Database \citep[EDD][]{jacobs_2009} presents a TRGB-based distance using two CMDs.\footnote{The data for \gal can be found at the following URL: \url{http://edd.ifa.hawaii.edu/get_cmd.php?pgc=50063}, which presents the visualizations associated with the ``CMDs/TRGB'' entry for \gal.}
The two CMDs are derived from data taken on HST via programs GO\,9492 (PI: Bresolin) and GO\,13691 (PI:Freedman, the CCHP field).
The preferred TRGB detection for the EDD is 29.08 mag.
The EDD uses either HSTPHOT or DOLPHOT for WFPC2 or ACS, respectively to produce CMDs to which the \citet{Makarov_2006} Maximum-Likelihood algorithm is applied and the \citetalias{rizzi_2007} zero-point is adopted.\footnote{This text is adapted from the EDD ``CMDs/TRGB'' description, which can be found on its main page, e.g., \url{http://edd.ifa.hawaii.edu/}}

\responsetoref{Given the data processing discrepancy identified by \citetalias{janglee_2017} in \citetalias{lee_2012}, the data analysis techniques for \citetalias{sakai_2004}, \citetalias{rizzi_2007}, and \citetalias{shappee_2011} were carefully reviewed to determine if the large range of distance moduli found in these works could be understood relative to data processing choices.} 
\responsetoref{ \citetalias{sakai_2004} follows the \citet{Hill_1998} photometry prescription, which does not employ a CTE correction, and adopts aperture corrections from distinct \hst observations, more specifically those corrections determined in observations of the dwarf galaxy Leo I presented in \citeauthor{Hill_1998}.
The differences between the more crowded \gal field as well as telescope ``breathing'' or other PSF modulations mean that these aperture corrections may not be well suited to the \gal data.
Thus, the \citetalias{sakai_2004} result, which is the largest TRGB-based distance modulus, could be attributed to these aspects of the data processing.}

\responsetoref{
On the other hand, \citetalias[][]{rizzi_2007} analyzed the \citetalias{sakai_2004} field using the HSTPHOT code, a precursor to DOLPhot that operates on similar underlying techniques in terms of ``native'' PSF modelling, application of CTE corrections, and determination of aperture corrections \citep{dolphin_2000,Dolphin_2000b}. 
The \citetalias{rizzi_2007} TRGB agrees with that of \citetalias{lee_2012} that used the WFPC2 module for the later DOLPhot code, which should operate under similar principles, but is still more distant than results from \hst$+$ACS at the $\sim$0.2~mag level (10\% in distance). 
That these latter two studies produce statistically indistinguishable results despite performing independent analyses, does suggest that there is something intrinsic to either this field or this dataset.   
We therefore suspect, as was discussed in \citetalias{janglee_2017}, that the depth of this WFPC2 pointing or something other characteristic about its location within \gal is the likely source of the ``long'' TRGB distances to \gal measured by \citetalias{sakai_2004}, \citetalias[][]{rizzi_2007}, and \citetalias{lee_2012}.}

\subsection{Detailed Comparison to JL17}

The TRGB detection of \citetalias{janglee_2017} of $F814W$ = 25.16 $\pm$ 0.035~mag, despite having similar statistical uncertainties, is 3-4$\sigma$ from the result in the CCHP field; even considering our total uncertainty, there is still a non-negligible difference between these results. 
However, this result uses the M101\_E field that was included in our archival analyses and, thus, we are able to do a detailed comparison to \citetalias{janglee_2017} and attempt to understand the origin of the discrepancy. 
Additional supporting descriptions and figures are given in \autoref{app:jl17_comp}. 

With the caveats of \autoref{sec:arc_trgb} in mind, we repeat the ASLF procedure for the M101\_E using our best estimation of the LF.
We selected the optical smoothing factor as the value that minimizes the combined uncertainty, which occurs at \sigmasmooth~=~0.05~mag.
The corresponding statistical and systematic uncertainties are \trgbobsvalstaterrROUNDED~mag and \trgbobsvalsyserrROUNDED~mag, respectively; we note, however, that owing to the complexity of modeling the non-RGB and non-AGB stellar populations contributing to the LF these likely represent lower limits on the uncertainties.  
We applied a color-cut to the CMD that is defined by the CCHP Field (M101\_A), but broadened commensurate with the larger photometric uncertainties for this field; \citetalias{janglee_2017} also apply a color cut in their analysis.
The result is a TRGB detection at \responsetoref{\tipf~$\pm$ \trgbobsvalstaterrROUNDED~mag}, though we note that the detection peak is asymmetric, but the asymmetry does not distort the result. 
Moreover, from \autoref{fig:smoothing}(v), both narrower and broader smoothing windows trend toward a measurement closer to 25.15~mag, albeit with larger uncertainties. 
The foreground extinction to this field is \aif~mag $\pm$ \eaif~mag such that the extinction corrected tip is \tipfcorr $\pm$ \tipfcorrerr ~(stat) $\pm$\trgbobsvalsyserrROUNDED~mag (sys).
This result sits in between the measurement from \citetalias{janglee_2017} and that from the CCHP field (M101\_A) and its uncertainties, taken together, are 1-$\sigma$ consistent with either result. 
From this, we conclude that the difference in the distance measurement between this work and \citetalias{janglee_2017} is dominated by the field choice, with differences in the underlying data processing or analysis playing a sub-dominant effect. 

\subsection{Comparison to Cepheid Distances}
There are numerous Cepheid distances to \gal and those distances published since the Key Project are compared in \autoref{fig:distcomp}b.
The distances were retrieved from NED and no homogenization has been applied owing to the complexities therein. 
We do note that the bulk of the studies use the Key Project distance modulus to the Large Magellanic Cloud of 18.50~mag \citep{freedman_2001}, which has shifted only by 0.01~mag in intervening years \citep[see discussion in][]{degrijs_2014} and thus the absolute scale of the distances has changed a small amount.
Thus, the differences between studies can be attributed to the exact parameters and wavelength of the Leavitt Law, treatment of stellar crowding in the photometry, assumptions regarding internal extinction, and assumptions regarding the metallicity term in the Leavitt Law.  

Since the \citetalias{lee_2012} compilation, additional Cepheid-distances have been measured by \citet{mager_2013}, \citet{Tully_2013}, \citet{nataf_2015}, and \citet{riess_2016}. 
The distance in \citet{nataf_2015} is not independent because it adopts data from \citetalias[][]{shappee_2011}, but performs a test of variant forms of the Milky Way interstellar extinction curve (with a total change $\Delta \mu$ = 0.06 mag) and, while the results are of great interest, we exclude them to avoid overemphasizing \citetalias[][]{shappee_2011} and complicating comparisons.\footnote{\textcolor{red}{In an earlier version of this manuscript posted to ArXiv the Nataf (2015) results were incorrectly presented. The authors apologize sincerely for this error and are thankful that it was corrected before publication.}}
Similar to the tabulation of \citetalias{lee_2012}, the distances to \gal via the Leavitt Law span 1~mag, with typical quoted uncertainties at the $\sim$0.1~mag level.

\subsection{Comparison of Pop~I and Pop~II Distances to \gal}

Using the four distances determined from TRGB methods, the weighted mean distance modulus is $\mu_{wt,mean}$ =29.11~$\pm$~0.02 mag and the unweighted mean distance modulus is $\mu_{mean}$ = 29.11~$\pm$~0.03 mag.
These values are plotted in \autoref{fig:distcomp}a as the vertical dashed and dot-dashed lines, respectively. 
We note that the``long distances'' measured in the \citetalias{sakai_2004}-field are excluded, as are those from from \citetalias{Tikhonov_2015} due to a lack of uncertainties (shown in grey in \autoref{fig:distcomp}a).
If we exclude \citetalias{janglee_2017} due to field choice, as previously discussed, then our mean value is determined using the distance modulus in this work and that from \citetalias{shappee_2011}, with a result of $\mu_{mean}$ = 29.08~mag $\pm$ 0.02 (uncertainty weighted).

The weighted mean of 13 distance moduli determined for \gal using the Leavitt Law is $\mu_{wt,mean}$ =~29.16~$\pm$~0.02 mag and the unweighted mean is $\mu_{mean}$ = 29.16~$\pm$~0.03 mag. 
These values are plotted in \autoref{fig:distcomp}b as the vertical dashed and dot-dashed lines, respectively.
From inspection of \autoref{fig:distcomp}b, the results before and after 2005 seem to cluster differently. 
If we limit to only the five distance moduli determined in 2005 and later, the weighted mean is $\mu_{wt,mean}$ = 29.12~$\pm$~0.03 mag and the unweighted mean is $\mu_{mean}$ = 29.11~$\pm$~0.05 mag. 

Thus, the ``mean'' distance to \gal from the Leavitt Law and the TRGB disagree at the $\sim$2$\sigma$-level, though the difference largely depends on what combination of measurements one chooses to compare. 
The difference is slightly surprising because the absolute scale, in either system, is set in the Large Magellanic Cloud, albeit the SH0ES program uses a set of objects to define the absolute scale \citep[see e.g.,][]{riess_2016}. 
We note that in \citet{riess_2016}, the calibration of SN2011fe in \gal is an outlier in the calibration relation (their figure 10) and the slightly closer distance measured by the TRGB resolves the discrepancy.

For context, we provide the ranges for distances determined from other standard candles, but note that none of these have comparable precision and accuracy as either the TRGB- or Cepheid-based scales.
Distance moduli determined from the \sn, itself, span a range from 28.86~mag to 29.38~mag.
Distance moduli determined from the Tully-Fisher relationship span a range from 27.07~mag to 29.62~mag and distance moduli determined from the planetary nebula luminosity function (PNLF) span a range from 29.36~mag to 29.42~mag. 
\responsetoref{Lastly, \citet{Carlsten_19a} presented distances using surface brightness fluctuations to individual satellite galaxies of \gal finding a mean (and median) distance of 6.5 $\pm$ 0.12 Mpc ($\mu$ = 29.10~mag) with a dispersion of 0.35 Mpc using the nine confirmed \gal satellites. 
Notably, the surface brightness fluctuation method is, itself, calibrated to the TRGB and galaxies of these low stellar masses typically have small quantities of dust \citep[the full method is described][]{Carlsten_19b}.}

\section{Summary} \label{sec:sum}

In this paper, we have determined the distance modulus to \gal using a carefully selected pointing \responsetoref{that is composed of a near pure Population II stars}.
The methods used in previous works in this series were converted into an end-to-end automated pipeline and produce results of comparable precision and accuracy \citepalias[][]{hatt_2017,jang_2018,hatt_2018a,hatt_2018b}.
We detect the TRGB at $m_{F814W}$=\trgbobsvalROUNDED~mag and determine the statistical and systematic uncertainties to be \trgbobsvalstaterrROUNDED~mag and \trgbobsvalsyserrROUNDED~mag, respectively. 
Dereddening the data and \responsetoref{using the final CCHP absolute magnitude calibration for the TRGB \citepalias{Freedman2019}}, we find a final true distance modulus of \truetrgbdmodwerr, which corresponds to a physical distance of \truetrgbdmodMpcwerr. 
The full error budget for this measurement is given in \autoref{tab_distance}. 
This distance is on the low-side of other distances relying on resolved stellar populations (\autoref{fig:distcomp}), but well within the NED range.
Unlike many of the literature methods, our technique is minimally affected from internal extinction and the impact from crowding is negligible.

We have used a set of archival images (\autoref{fig:pointingmap}a and \autoref{tab:datasets}) to demonstrate that proper selection of a field suitable for TRGB measurement is required to reach high {\it accuracy} -- even for \gal, one of the most nearby \sn host galaxies.
We demonstrated that while the TRGB discontinuity is visible by eye at the level detected in our CCHP field (e.g., \autoref{fig:cmds}), the edge detection algorithm is easily confused due to a combination of multiple stellar populations \responsetoref{and multiple structural components.
We caution that application of the TRGB in these scenarios will have large uncertainties that may not be encapsulated by commonly applied methods. 
Though we are able to show that the uncertainty as measured from the width of the response function does provide a meaningful uncertainty measure in the cases of large effective-smoothing.}
Additionally, the lower signal-to-noise produces larger magnitude and color uncertainties that amplify confusion by smearing the populations in the luminosity function.
We show in \autoref{fig:pointingmap_all}b that the surface brightness profiles provide insight into the selection of an appropriate pointing free from these contaminating populations, with the combination of all-sky-survey depth \emph{GALEX} and \emph{WISE} imaging being of sufficient quality.  
If available in the literature, HI maps can also inform field choice to limit the impact from dust.

\acknowledgments

We warmly thank the anonymous referee for their careful attention to the manuscript that has improved work presented here. 
We thank Peter Stetson for a copy of the \textsc{daophot} family of programs as well as his helpful interactions on its use for our science case. 
RLB thanks Sean Johnson for engaging discussions on gas and dust in the CGM, Saurabh Jha for numerous insights that made their way into this work, and Adam Riess for a helpful discussion of crowding-based systematics. 
In version of this manuscript posted to ArXiv, the \citet{nataf_2015} results were incorrectly presented. The authors apologize sincerely for this error and are thankful that it was corrected before publication.
Support for this work was provided by NASA through Hubble Fellowship grant \#51386.01 awarded to R.L.B.by the Space Telescope Science Institute, which is operated by the Association of  Universities for Research in Astronomy, Inc., for NASA, under contract NAS 5-26555.

Support for program \#13691 was provided by NASA through a grant from the Space Telescope Science Institute, which is operated by the Association of Universities for Research in Astronomy, Inc., under NASA contract NAS 5-26555.
MGL was supported by the National Research Foundation grant funded by the Korean Government  (NRF-2017R1A2B4004632).
Some of the data presented in this paper were obtained from the Mikulski Archive for Space Telescopes (MAST). STScI is operated by the Association of Universities for Research in Astronomy, Inc., under NASA contract NAS5-26555. 

This research has made use of the NASA/IPAC Extragalactic Database (NED) which is operated by the Jet Propulsion Laboratory, California Institute of Technology, under contract with the National Aeronautics and Space Administration.

This research has made use of the NASA/IPAC Infrared Science Archive (IRSA), which is operated by the Jet Propulsion Laboratory, California Institute of Technology, under contract with the National Aeronautics and Space Administration. 

This work made used of THINGS, `{\it The HI nearby Galaxy Survey}' \citep{walter_2008}. Available: \url{ http://www.mpia.de/THINGS/Overview.html}


This publication makes use of data products from the Wide-field Infrared Survey Explorer, which is a joint project of the University of California, Los Angeles, and the Jet Propulsion Laboratory/California Institute of Technology, funded by the National Aeronautics and Space Administration.


Based on observations made with the NASA Galaxy Evolution Explorer. GALEX is operated for NASA by the California Institute of Technology under NASA contract NAS5-98034.

\vspace{5mm}
\facilities{HST (ACS/WFC)}
\software{DAOPHOT \citep{stetson_1987}, ALLFRAME \citep{stetson_1994}}


\begin{thebibliography}{}
\expandafter\ifx\csname natexlab\endcsname\relax\def\natexlab#1{#1}\fi
\providecommand{\url}[1]{\href{#1}{#1}}

\bibitem[{{Alves} \& {Cook}(1995)}]{alves_1995}
{Alves}, D.~R., \& {Cook}, K.~H. 1995, \aj, 110, 192

\bibitem[{{Anand} {et~al.}(2018){Anand}, {Rizzi}, \& {Tully}}]{anand_2018}
{Anand}, G.~S., {Rizzi}, L., \& {Tully}, R.~B. 2018, \aj, 156, 105

\bibitem[{{Beaton} {et~al.}(2016){Beaton}, {Freedman}, {Madore}, {Bono},
  {Carlson}, {Clementini}, {Durbin}, {Garofalo}, {Hatt}, {Jang}, {Kollmeier},
  {Lee}, {Monson}, {Rich}, {Scowcroft}, {Seibert}, {Sturch}, \&
  {Yang}}]{beaton_2016}
{Beaton}, R.~L., {Freedman}, W.~L., {Madore}, B.~F., {et~al.} 2016, \apj, 832,
  210

\bibitem[{{Beaton} {et~al.}(2018){Beaton}, {Bono}, {Braga}, {Dall'Ora},
  {Fiorentino}, {Jang}, {Mart{\'{\i}}nez-V{\'a}zquez}, {Matsunaga}, {Monelli},
  {Neeley}, \& {Salaris}}]{beaton_2018}
{Beaton}, R.~L., {Bono}, G., {Braga}, V.~F., {et~al.} 2018, \ssr, 214, 113

\bibitem[{{Bohlin}(2016)}]{boh16}
{Bohlin}, R.~C. 2016, \aj, 152, 60

\bibitem[{{Calzetti}(2013)}]{prop13364}
{Calzetti}, D. 2013, {LEGUS: Legacy ExtraGalactic UV Survey}, HST Proposal, ,

\bibitem[{{Cardelli} {et~al.}(1989){Cardelli}, {Clayton}, \&
  {Mathis}}]{cardelli_1989}
{Cardelli}, J.~A., {Clayton}, G.~C., \& {Mathis}, J.~S. 1989, \apj, 345, 245

\bibitem[{{Carlsten} {et~al.}(2019{\natexlab{a}}){Carlsten}, {Beaton}, {Greco},
  \& {Greene}}]{Carlsten_19a}
{Carlsten}, S.~G., {Beaton}, R.~L., {Greco}, J.~P., \& {Greene}, J.~E.
  2019{\natexlab{a}}, \apjl, 878, L16

\bibitem[{{Carlsten} {et~al.}(2019{\natexlab{b}}){Carlsten}, {Beaton}, {Greco},
  \& {Greene}}]{Carlsten_19b}
---. 2019{\natexlab{b}}, \apj, 879, 13

\bibitem[{{Carraro}(2015)}]{carraro_2015}
{Carraro}, G. 2015, Boletin de la Asociacion Argentina de Astronomia La Plata
  Argentina, 57, 138

\bibitem[{{Cook} {et~al.}(1986){Cook}, {Aaronson}, \&
  {Illingworth}}]{cook_1986}
{Cook}, K.~H., {Aaronson}, M., \& {Illingworth}, G. 1986, \apjl, 301, L45

\bibitem[{{Cook} {et~al.}(1989){Cook}, {Aaronson}, \&
  {Illingworth}}]{cook_1989}
{Cook}, K.~H., {Aaronson}, M., \& {Illingworth}, G. 1989, in \baas, Vol.~21,
  Bulletin of the American Astronomical Society, 719

\bibitem[{{de Grijs} {et~al.}(2014){de Grijs}, {Wicker}, \&
  {Bono}}]{degrijs_2014}
{de Grijs}, R., {Wicker}, J.~E., \& {Bono}, G. 2014, \aj, 147, 122

\bibitem[{{Dolphin}(2000{\natexlab{a}})}]{dolphin_2000}
{Dolphin}, A.~E. 2000{\natexlab{a}}, \pasp, 112, 1383

\bibitem[{{Dolphin}(2000{\natexlab{b}})}]{Dolphin_2000b}
---. 2000{\natexlab{b}}, \pasp, 112, 1397

\bibitem[{{Ferrarese} {et~al.}(2000){Ferrarese}, {Mould}, {Kennicutt},
  {Huchra}, {Ford}, {Freedman}, {Stetson}, {Madore}, {Sakai}, {Gibson},
  {Graham}, {Hughes}, {Illingworth}, {Kelson}, {Macri}, {Sebo}, \&
  {Silbermann}}]{ferrarese_2000}
{Ferrarese}, L., {Mould}, J.~R., {Kennicutt}, Jr., R.~C., {et~al.} 2000, \apj,
  529, 745

\bibitem[{{Freedman}(2014)}]{prop13691}
{Freedman}, W. 2014, {CHP-II: The Carnegie Hubble Program to Measure Ho to 3\%
  Using Population II}, HST Proposal, ,

\bibitem[{{Freedman}(2017)}]{freedman_2017}
{Freedman}, W.~L. 2017, Nature Astronomy, 1, 0121

\bibitem[{{Freedman} \& {Madore}(2010)}]{freedman_2010}
{Freedman}, W.~L., \& {Madore}, B.~F. 2010, \araa, 48, 673

\bibitem[{{Freedman} {et~al.}(2012){Freedman}, {Madore}, {Scowcroft}, {Burns},
  {Monson}, {Persson}, {Seibert}, \& {Rigby}}]{freedman_2012}
{Freedman}, W.~L., {Madore}, B.~F., {Scowcroft}, V., {et~al.} 2012, \apj, 758,
  24

\bibitem[{{Freedman} {et~al.}(2001){Freedman}, {Madore}, {Gibson}, {Ferrarese},
  {Kelson}, {Sakai}, {Mould}, {Kennicutt}, {Ford}, {Graham}, {Huchra},
  {Hughes}, {Illingworth}, {Macri}, \& {Stetson}}]{freedman_2001}
{Freedman}, W.~L., {Madore}, B.~F., {Gibson}, B.~K., {et~al.} 2001, \apj, 553,
  47

\bibitem[{{Freedman} {et~al.}(2019){Freedman}, {Madore}, {Hatt}, {Hoyt},
  {Jang}, {Beaton}, {Burns}, {Lee}, {Monson}, {Neeley}, {Phillips}, {Rich}, \&
  {Seibert}}]{Freedman2019}
{Freedman}, W.~L., {Madore}, B.~F., {Hatt}, D., {et~al.} 2019, arXiv e-prints,
  arXiv:1907.05922

\bibitem[{{Gil de Paz} {et~al.}(2007){Gil de Paz}, {Boissier}, {Madore},
  {Seibert}, {Joe}, {Boselli}, {Wyder}, {Thilker}, {Bianchi}, {Rey}, {Rich},
  {Barlow}, {Conrow}, {Forster}, {Friedman}, {Martin}, {Morrissey}, {Neff},
  {Schiminovich}, {Small}, {Donas}, {Heckman}, {Lee}, {Milliard}, {Szalay}, \&
  {Yi}}]{gildepaz_2007}
{Gil de Paz}, A., {Boissier}, S., {Madore}, B.~F., {et~al.} 2007, \apjs, 173,
  185

\bibitem[{{Hatt} {et~al.}(2017){Hatt}, {Beaton}, {Freedman}, {Madore}, {Jang},
  {Hoyt}, {Lee}, {Monson}, {Rich}, {Scowcroft}, \& {Seibert}}]{hatt_2017}
{Hatt}, D., {Beaton}, R.~L., {Freedman}, W.~L., {et~al.} 2017, \apj, 845, 146

\bibitem[{{Hatt} {et~al.}(2018{\natexlab{a}}){Hatt}, {Freedman}, {Madore},
  {Beaton}, {Hoyt}, {Jang}, {Lee}, {Monson}, {Rich}, {Scowcroft}, \&
  {Seibert}}]{hatt_2018a}
{Hatt}, D., {Freedman}, W.~L., {Madore}, B.~F., {et~al.} 2018{\natexlab{a}},
  \apj, 861, 104

\bibitem[{{Hatt} {et~al.}(2018{\natexlab{b}}){Hatt}, {Freedman}, {Madore},
  {Jang}, {Beaton}, {Hoyt}, {Lee}, {Monson}, {Rich}, {Scowcroft}, \&
  {Seibert}}]{hatt_2018b}
---. 2018{\natexlab{b}}, \apj, 866, 145

\bibitem[{{Hill} {et~al.}(1998){Hill}, {Ferrarese}, {Stetson}, {Saha},
  {Freedman}, {Graham}, {Hoessel}, {Han}, {Huchra}, \& {Hughes}}]{Hill_1998}
{Hill}, R.~J., {Ferrarese}, L., {Stetson}, P.~B., {et~al.} 1998, \apj, 496, 648

\bibitem[{{Hoyt} {et~al.}(2018){Hoyt}, {Freedman}, {Madore}, {Seibert},
  {Beaton}, {Hatt}, {Jang}, {Lee}, {Monson}, \& {Rich}}]{hoyt_2018}
{Hoyt}, T.~J., {Freedman}, W.~L., {Madore}, B.~F., {et~al.} 2018, \apj, 858, 12

\bibitem[{{Hoyt} {et~al.}(2019){Hoyt}, {Freedman}, {Madore}, {Beaton}, {Hatt},
  {Jang}, {Lee}, {Monson}, {Neeley}, {Rich}, \& {Mager}}]{Hoyt2019}
---. 2019, arXiv e-prints, arXiv:1907.05891

\bibitem[{{Jacobs} {et~al.}(2009){Jacobs}, {Rizzi}, {Tully}, {Shaya},
  {Makarov}, \& {Makarova}}]{jacobs_2009}
{Jacobs}, B.~A., {Rizzi}, L., {Tully}, R.~B., {et~al.} 2009, \aj, 138, 332

\bibitem[{{Jang} \& {Lee}(2017{\natexlab{a}})}]{janglee_2017b}
{Jang}, I.~S., \& {Lee}, M.~G. 2017{\natexlab{a}}, \apj, 836, 74

\bibitem[{{Jang} \& {Lee}(2017{\natexlab{b}})}]{janglee_2017}
---. 2017{\natexlab{b}}, \apj, 835, 28

\bibitem[{{Jang} {et~al.}(2018){Jang}, {Hatt}, {Beaton}, {Lee}, {Freedman},
  {Madore}, {Hoyt}, {Monson}, {Rich}, {Scowcroft}, \& {Seibert}}]{jang_2018}
{Jang}, I.~S., {Hatt}, D., {Beaton}, R.~L., {et~al.} 2018, \apj, 852, 60

\bibitem[{{Kelson} {et~al.}(1996){Kelson}, {Illingworth}, {Freedman}, {Graham},
  {Hill}, {Madore}, {Saha}, {Stetson}, {Kennicutt}, {Mould}, {Hughes},
  {Ferrarese}, {Phelps}, {Turner}, {Cook}, {Ford}, {Hoessel}, \&
  {Huchra}}]{kelson_1996}
{Kelson}, D.~D., {Illingworth}, G.~D., {Freedman}, W.~F., {et~al.} 1996, \apj,
  463, 26

\bibitem[{{Kennicutt} {et~al.}(1998){Kennicutt}, {Stetson}, {Saha}, {Kelson},
  {Rawson}, {Sakai}, {Madore}, {Mould}, {Freedman}, {Bresolin}, {Ferrarese},
  {Ford}, {Gibson}, {Graham}, {Han}, {Harding}, {Hoessel}, {Huchra}, {Hughes},
  {Illingworth}, {Macri}, {Phelps}, {Silbermann}, {Turner}, \&
  {Wood}}]{Kennicutt_1998}
{Kennicutt}, Jr., R.~C., {Stetson}, P.~B., {Saha}, A., {et~al.} 1998, \apj,
  498, 181

\bibitem[{{Krist} {et~al.}(2011){Krist}, {Hook}, \& {Stoehr}}]{krist_2011}
{Krist}, J.~E., {Hook}, R.~N., \& {Stoehr}, F. 2011, in \procspie, Vol. 8127,
  Optical Modeling and Performance Predictions V, 81270J

\bibitem[{{Lee} {et~al.}(1993){Lee}, {Freedman}, \& {Madore}}]{lee_1993}
{Lee}, M.~G., {Freedman}, W.~L., \& {Madore}, B.~F. 1993, \apj, 417, 553

\bibitem[{{Lee} \& {Jang}(2012)}]{lee_2012}
{Lee}, M.~G., \& {Jang}, I.~S. 2012, \apjl, 760, L14

\bibitem[{{Mack} {et~al.}(2007){Mack}, {Gilliland}, {Anderson}, \&
  {Sirianni}}]{mack_2007}
{Mack}, J., {Gilliland}, R.~L., {Anderson}, J., \& {Sirianni}, M. 2007, {WFC
  Zeropoints at -80C}, Tech. rep.

\bibitem[{{Macri} {et~al.}(2001){Macri}, {Calzetti}, {Freedman}, {Gibson},
  {Graham}, {Huchra}, {Hughes}, {Madore}, {Mould}, {Persson}, \&
  {Stetson}}]{macri_2001}
{Macri}, L.~M., {Calzetti}, D., {Freedman}, W.~L., {et~al.} 2001, \apj, 549,
  721

\bibitem[{{Madore} \& {Freedman}(1995)}]{madore_1995}
{Madore}, B.~F., \& {Freedman}, W.~L. 1995, \aj, 109, 1645

\bibitem[{{Madore} \& {Freedman}(1999)}]{madore_1999}
{Madore}, B.~F., \& {Freedman}, W.~L. 1999, in Astronomical Society of the
  Pacific Conference Series, Vol. 167, Harmonizing Cosmic Distance Scales in a
  Post-HIPPARCOS Era, ed. D.~{Egret} \& A.~{Heck}, 161--174

\bibitem[{{Madore} {et~al.}(2009){Madore}, {Mager}, \&
  {Freedman}}]{Madore_2009}
{Madore}, B.~F., {Mager}, V., \& {Freedman}, W.~L. 2009, \apj, 690, 389

\bibitem[{{Madore} {et~al.}(2018){Madore}, {Freedman}, {Hatt}, {Hoyt},
  {Monson}, {Beaton}, {Rich}, {Jang}, {Lee}, {Scowcroft}, \&
  {Seibert}}]{madore_2018}
{Madore}, B.~F., {Freedman}, W.~L., {Hatt}, D., {et~al.} 2018, \apj, 858, 11

\bibitem[{{Mager} {et~al.}(2008){Mager}, {Madore}, \& {Freedman}}]{Mager_2008}
{Mager}, V.~A., {Madore}, B.~F., \& {Freedman}, W.~L. 2008, \apj, 689, 721

\bibitem[{{Mager} {et~al.}(2013){Mager}, {Madore}, \& {Freedman}}]{mager_2013}
---. 2013, \apj, 777, 79

\bibitem[{{Makarov} {et~al.}(2006){Makarov}, {Makarova}, {Rizzi}, {Tully},
  {Dolphin}, {Sakai}, \& {Shaya}}]{Makarov_2006}
{Makarov}, D., {Makarova}, L., {Rizzi}, L., {et~al.} 2006, \aj, 132, 2729

\bibitem[{{McQuinn} {et~al.}(2016{\natexlab{a}}){McQuinn}, {Skillman},
  {Dolphin}, {Berg}, \& {Kennicutt}}]{McQuinn_2016a}
{McQuinn}, K. B.~W., {Skillman}, E.~D., {Dolphin}, A.~E., {Berg}, D., \&
  {Kennicutt}, R. 2016{\natexlab{a}}, \apj, 826, 21

\bibitem[{{McQuinn} {et~al.}(2016{\natexlab{b}}){McQuinn}, {Skillman},
  {Dolphin}, {Berg}, \& {Kennicutt}}]{McQuinn_2016b}
---. 2016{\natexlab{b}}, \aj, 152, 144

\bibitem[{{McQuinn} {et~al.}(2017){McQuinn}, {Skillman}, {Dolphin}, {Berg}, \&
  {Kennicutt}}]{McQuinn_2017}
---. 2017, \aj, 154, 51

\bibitem[{{M{\'e}ndez} {et~al.}(2002){M{\'e}ndez}, {Davis}, {Moustakas},
  {Newman}, {Madore}, \& {Freedman}}]{mendez_2002}
{M{\'e}ndez}, B., {Davis}, M., {Moustakas}, J., {et~al.} 2002, \aj, 124, 213

\bibitem[{{Monson} {et~al.}(2017){Monson}, {Beaton}, {Scowcroft}, {Freedman},
  {Madore}, {Rich}, {Seibert}, {Kollmeier}, \& {Clementini}}]{monson_2017}
{Monson}, A.~J., {Beaton}, R.~L., {Scowcroft}, V., {et~al.} 2017, \aj, 153, 96

\bibitem[{{Nataf}(2015)}]{nataf_2015}
{Nataf}, D.~M. 2015, \mnras, 449, 1171

\bibitem[{{Newman} {et~al.}(2001){Newman}, {Ferrarese}, {Stetson}, {Maoz},
  {Zepf}, {Davis}, {Freedman}, \& {Madore}}]{newman_2001}
{Newman}, J.~A., {Ferrarese}, L., {Stetson}, P.~B., {et~al.} 2001, \apj, 553,
  562

\bibitem[{{Nugent} {et~al.}(2011{\natexlab{a}}){Nugent}, {Sullivan}, {Bersier},
  {Howell}, {Thomas}, \& {James}}]{11fe_discovery}
{Nugent}, P., {Sullivan}, M., {Bersier}, D., {et~al.} 2011{\natexlab{a}}, The
  Astronomer's Telegram, 3581

\bibitem[{{Nugent} {et~al.}(2011{\natexlab{b}}){Nugent}, {Sullivan}, {Cenko},
  {Thomas}, {Kasen}, {Howell}, {Bersier}, {Bloom}, {Kulkarni}, {Kandrashoff},
  {Filippenko}, {Silverman}, {Marcy}, {Howard}, {Isaacson}, {Maguire},
  {Suzuki}, {Tarlton}, {Pan}, {Bildsten}, {Fulton}, {Parrent}, {Sand},
  {Podsiadlowski}, {Bianco}, {Dilday}, {Graham}, {Lyman}, {James}, {Kasliwal},
  {Law}, {Quimby}, {Hook}, {Walker}, {Mazzali}, {Pian}, {Ofek}, {Gal-Yam}, \&
  {Poznanski}}]{nugent_2011}
{Nugent}, P.~E., {Sullivan}, M., {Cenko}, S.~B., {et~al.} 2011{\natexlab{b}},
  \nat, 480, 344

\bibitem[{{Overbye}(1991)}]{lonely_hearts}
{Overbye}, D. 1991, {Lonely hearts of the cosmos. The scientific quest for the
  secret of the universe.}

\bibitem[{{Paturel} {et~al.}(2002){Paturel}, {Teerikorpi}, {Theureau},
  {Fouqu{\'e}}, {Musella}, \& {Terry}}]{paturel_2002}
{Paturel}, G., {Teerikorpi}, P., {Theureau}, G., {et~al.} 2002, \aap, 389, 19

\bibitem[{{Peek} {et~al.}(2015){Peek}, {M{\'e}nard}, \& {Corrales}}]{peek_2015}
{Peek}, J.~E.~G., {M{\'e}nard}, B., \& {Corrales}, L. 2015, \apj, 813, 7

\bibitem[{{Persson} {et~al.}(2004){Persson}, {Madore}, {Krzemi{\'n}ski},
  {Freedman}, {Roth}, \& {Murphy}}]{persson_2004}
{Persson}, S.~E., {Madore}, B.~F., {Krzemi{\'n}ski}, W., {et~al.} 2004, \aj,
  128, 2239

\bibitem[{{Riess} {et~al.}(2019){Riess}, {Casertano}, {Yuan}, {Macri}, \&
  {Scolnic}}]{riess_2019}
{Riess}, A.~G., {Casertano}, S., {Yuan}, W., {Macri}, L.~M., \& {Scolnic}, D.
  2019, \apj, 876, 85

\bibitem[{{Riess} {et~al.}(2011){Riess}, {Macri}, {Casertano}, {Lampeitl},
  {Ferguson}, {Filippenko}, {Jha}, {Li}, \& {Chornock}}]{riess_2011}
{Riess}, A.~G., {Macri}, L., {Casertano}, S., {et~al.} 2011, \apj, 730, 119

\bibitem[{{Riess} {et~al.}(2016){Riess}, {Macri}, {Hoffmann}, {Scolnic},
  {Casertano}, {Filippenko}, {Tucker}, {Reid}, {Jones}, {Silverman},
  {Chornock}, {Challis}, {Yuan}, {Brown}, \& {Foley}}]{riess_2016}
{Riess}, A.~G., {Macri}, L.~M., {Hoffmann}, S.~L., {et~al.} 2016, \apj, 826, 56

\bibitem[{{Riess} {et~al.}(2018){Riess}, {Casertano}, {Yuan}, {Macri},
  {Anderson}, {Mackenty}, {Bowers}, {Clubb}, {Filippenko}, {Jones}, \&
  {Tucker}}]{riess_2018}
{Riess}, A.~G., {Casertano}, S., {Yuan}, W., {et~al.} 2018, ArXiv e-prints,
  arXiv:1801.01120

\bibitem[{{Rizzi} {et~al.}(2007){Rizzi}, {Tully}, {Makarov}, {Makarova},
  {Dolphin}, {Sakai}, \& {Shaya}}]{rizzi_2007}
{Rizzi}, L., {Tully}, R.~B., {Makarov}, D., {et~al.} 2007, \apj, 661, 815

\bibitem[{{Saha} {et~al.}(2006){Saha}, {Thim}, {Tammann}, {Reindl}, \&
  {Sandage}}]{saha_2006}
{Saha}, A., {Thim}, F., {Tammann}, G.~A., {Reindl}, B., \& {Sandage}, A. 2006,
  \apjs, 165, 108

\bibitem[{{Sakai} {et~al.}(2004){Sakai}, {Ferrarese}, {Kennicutt}, \&
  {Saha}}]{sakai_2004}
{Sakai}, S., {Ferrarese}, L., {Kennicutt}, Jr., R.~C., \& {Saha}, A. 2004,
  \apj, 608, 42

\bibitem[{{Sandage} \& {Tammann}(1974)}]{sandage_1974}
{Sandage}, A., \& {Tammann}, G.~A. 1974, \apj, 194, 223

\bibitem[{{Schlafly} \& {Finkbeiner}(2011)}]{sch11}
{Schlafly}, E.~F., \& {Finkbeiner}, D.~P. 2011, \apj, 737, 103

\bibitem[{{Schlegel} {et~al.}(1998){Schlegel}, {Finkbeiner}, \&
  {Davis}}]{schl98}
{Schlegel}, D.~J., {Finkbeiner}, D.~P., \& {Davis}, M. 1998, \apj, 500, 525

\bibitem[{{Serenelli} {et~al.}(2017){Serenelli}, {Weiss}, {Cassisi}, {Salaris},
  \& {Pietrinferni}}]{serenelli_2017}
{Serenelli}, A., {Weiss}, A., {Cassisi}, S., {Salaris}, M., \& {Pietrinferni},
  A. 2017, \aap, 606, A33

\bibitem[{{Shappee}(2014)}]{prop13737}
{Shappee}, B. 2014, {Whimper of a Bang: Documenting the Final Days of the
  Nearby Type Ia Supernova 2011fe}, HST Proposal, ,

\bibitem[{{Shappee}(2015)}]{prop14166}
---. 2015, {Whimper of a Bang: Documenting the Final Days of the Nearby Type Ia
  Supernova 2011fe}, HST Proposal, ,

\bibitem[{{Shappee} \& {Stanek}(2011)}]{shappee_2011}
{Shappee}, B.~J., \& {Stanek}, K.~Z. 2011, \apj, 733, 124

\bibitem[{{Shappee} {et~al.}(2016){Shappee}, {Stanek}, {Kochanek}, \&
  {Garnavich}}]{shappee_2016}
{Shappee}, B.~J., {Stanek}, K.~Z., {Kochanek}, C.~S., \& {Garnavich}, P.~M.
  2016, ArXiv e-prints, arXiv:1608.01155

\bibitem[{{Sirianni} {et~al.}(2005){Sirianni}, {Jee}, {Benitez}, {Blakeslee},
  {Martel}, {Meurer}, {Clampin}, {De Marchi}, {Ford}, {Gilliland}, {Hartig},
  {Illingworth}, {Mack}, \& {McCann}}]{sirianni_2005}
{Sirianni}, M., {Jee}, M.~J., {Benitez}, N., {et~al.} 2005, \pasp, 117, 1049

\bibitem[{{Steer} {et~al.}(2017){Steer}, {Madore}, {Mazzarella}, {Schmitz},
  {Corwin}, {Chan}, {Ebert}, {Helou}, {Baker}, {Chen}, {Frayer}, {Jacobson},
  {Lo}, {Ogle}, {Pevunova}, \& {Terek}}]{steer_2017}
{Steer}, I., {Madore}, B.~F., {Mazzarella}, J.~M., {et~al.} 2017, \aj, 153, 37

\bibitem[{{Stetson}(1987)}]{stetson_1987}
{Stetson}, P.~B. 1987, \pasp, 99, 191

\bibitem[{{Stetson}(1990)}]{stetson_1990}
---. 1990, \pasp, 102, 932

\bibitem[{{Stetson}(1994)}]{stetson_1994}
---. 1994, \pasp, 106, 250

\bibitem[{{Stetson} \& {Harris}(1988)}]{stetson_1988}
{Stetson}, P.~B., \& {Harris}, W.~E. 1988, \aj, 96, 909

\bibitem[{{Stetson} {et~al.}(1998){Stetson}, {Saha}, {Ferrarese}, {Rawson},
  {Ford}, {Freedman}, {Gibson}, {Graham}, {Harding}, {Han}, {Hill}, {Hoessel},
  {Huchra}, {Hughes}, {Illingworth}, {Kelson}, {Kennicutt}, {Madore}, {Mould},
  {Phelps}, {Sakai}, {Silbermann}, \& {Turner}}]{stetson_1998}
{Stetson}, P.~B., {Saha}, A., {Ferrarese}, L., {et~al.} 1998, \apj, 508, 491

\bibitem[{{Tikhonov} {et~al.}(2015){Tikhonov}, {Lebedev}, \&
  {Galazutdinova}}]{Tikhonov_2015}
{Tikhonov}, N.~A., {Lebedev}, V.~S., \& {Galazutdinova}, O.~A. 2015, Astronomy
  Letters, 41, 239

\bibitem[{{Tully} {et~al.}(2013){Tully}, {Courtois}, {Dolphin}, {Fisher},
  {H{\'e}raudeau}, {Jacobs}, {Karachentsev}, {Makarov}, {Makarova},
  {Mitronova}, {Rizzi}, {Shaya}, {Sorce}, \& {Wu}}]{Tully_2013}
{Tully}, R.~B., {Courtois}, H.~M., {Dolphin}, A.~E., {et~al.} 2013, \aj, 146,
  86

\bibitem[{{Walter} {et~al.}(2008){Walter}, {Brinks}, {de Blok}, {Bigiel},
  {Kennicutt}, {Thornley}, \& {Leroy}}]{walter_2008}
{Walter}, F., {Brinks}, E., {de Blok}, W.~J.~G., {et~al.} 2008, \aj, 136, 2563

\bibitem[{{Willick} \& {Batra}(2001)}]{Willick_2001}
{Willick}, J.~A., \& {Batra}, P. 2001, \apj, 548, 564

\bibitem[{{Zaritsky}(1994)}]{zaritsky_1994}
{Zaritsky}, D. 1994, \aj, 108, 1619

\end{thebibliography}


\appendix

\section{Supplementary Information}\label{app}
In this Appendix, we provide additional information regarding the photometry and analyses undertaken in the main text for the benefit of the reader and for reproducibility of the results. 

\subsection{Photometric Terms for Each Field}
\autoref{tab:calibration} presents the time dependent zeropoint (ZP), correction to an infinite aperture (APInf), and aperture core correction (ApCore$_{c1}$ for chip1 and ApCore$_{c2}$ for chip2) used for each of the \hst+ACS pointings in \gal. 
The STSci provided values (ZP and APInf) are sometimes adjusted retroactively at the $\sim$1\% level and these changes are not negligible for the precisions quoted here. 
The APCore correction in \autoref{tab:calibration} is the mean value for the series of frames to give a sense of the correction.
The values used for the photometry in the main text are applied on a frame-by-frame basis before computing the mean of the frame-by-frame instrumental magnitudes. 

\subsection{Image Quality and Magnitude Uncertainty Cuts for Each Field}

The photometry used in the main text is restricted using the photometric uncertainty ($\sigma_{\rm F814W}$), the sharpness parameter (sharp$_{\rm F814W}$), and the chi parameter ($\chi_{\rm F814W}$) that are reported by {\sc daophot}. 
We use a set of functions to fit the form of the distributions for these parameters as a function of magnitude and define likely stellar sources to be those that pass all three tests. 
The functions follow the following forms:

\begin{equation}
\sigma_{\rm F814W} < 0.03 + 0.003\times e^{m-m_{\rm \sigma}},
\end{equation}

\begin{equation}
\abs{sharp_{\rm F814W}} < 0.10 + 0.0075\times e^{m-m_{\rm sharp}}, 
\end{equation}

\begin{equation}
\chi_{\rm F814W} < 2.0 + e^{m-m_{\rm \chi}}, 
\end{equation}

\noindent where the $m_{\rm \sigma}$, $m_{\rm sharp}$, and $m_{\rm \chi}$ are determined for each field individually and generally scale with signal-to-noise. The value for each parameter are given for each field in \autoref{tab:calibration}. 

\begin{figure*} 
  \centering
    \includegraphics[width=0.48\columnwidth]{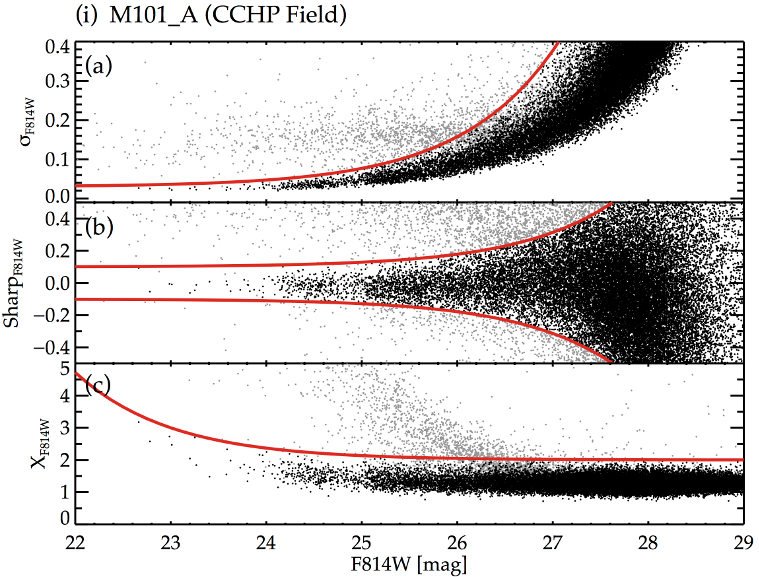} \\ 
    \includegraphics[width=0.48\columnwidth]{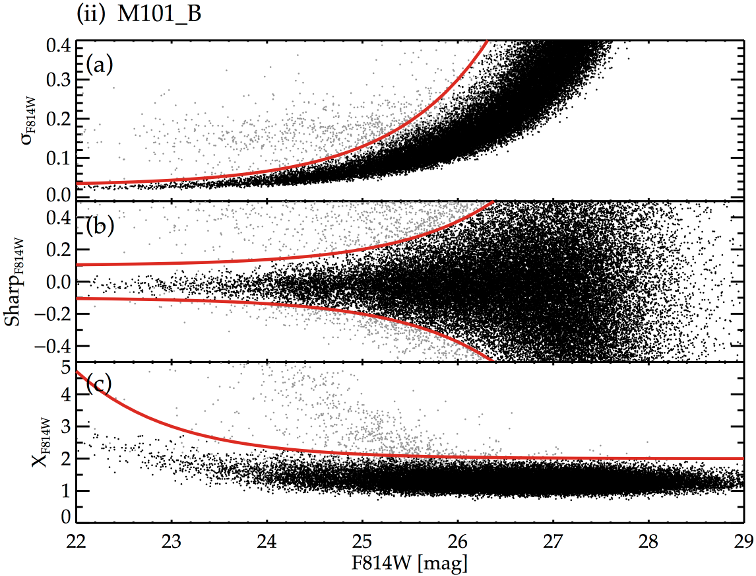}
    \includegraphics[width=0.48\columnwidth]{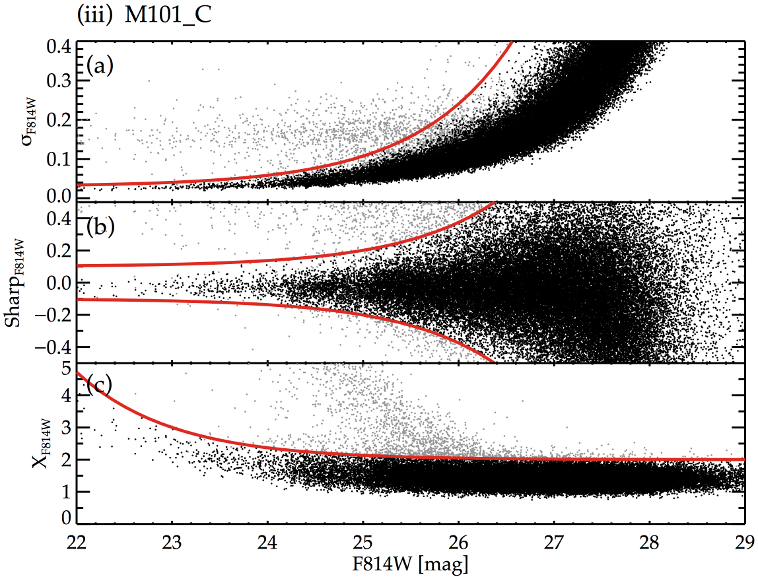}
    \includegraphics[width=0.48\columnwidth]{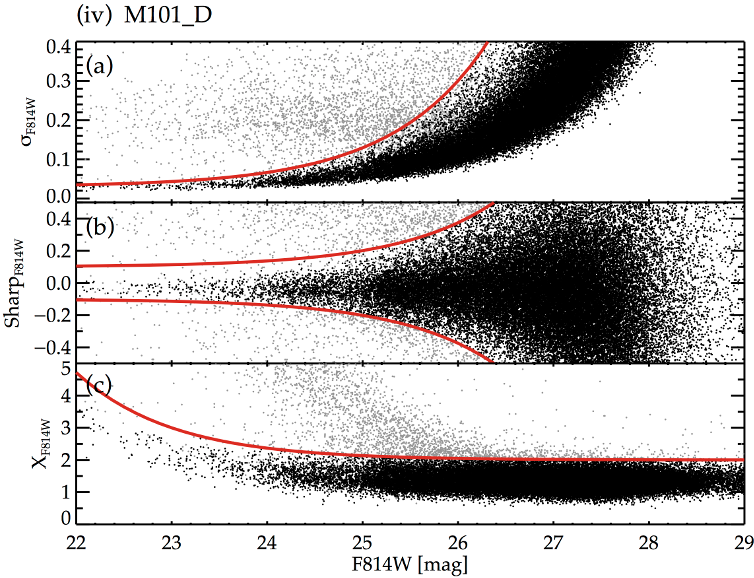}
    \includegraphics[width=0.48\columnwidth]{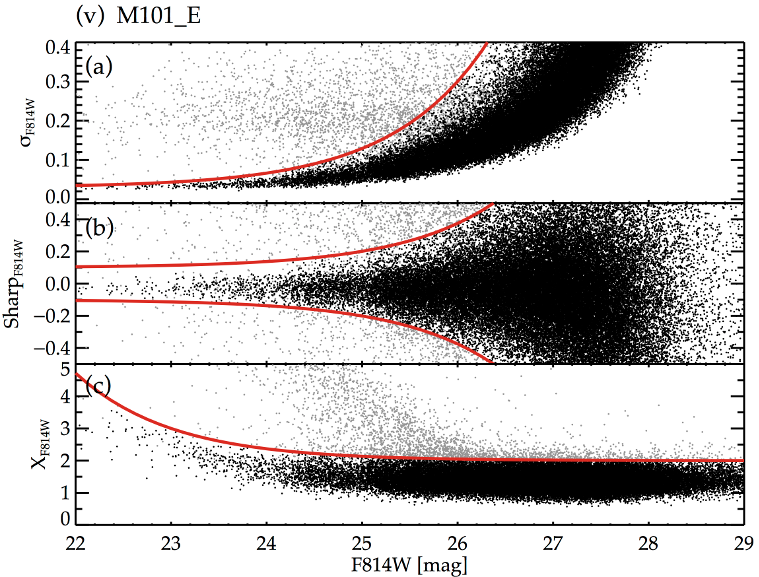}
  \caption{\label{fig:phot_qual_all} 
  Photometry quality diagrams for (i) M101\_A (CCHP field) and the four archival fields (ii) M101\_B, (iii) M101\_C, (iv) M101\_D, and (v) M101\_E. The sub-panels demonstrate (a) the photometric uncertainty and image quality metrics (b) sharpness parameter and (c) $\chi$ as a function of the F814W magnitude. The restrictions used on the photometry are shown in red with the sources passing all of the restrictions in black and the sources not passing any one criterion in gray. Generally, the archival fields are slightly shallower than our CCHP pointing, but are similar in signal-to-noise as observations in other \sn-hosts in the CCHP program.}
\end{figure*}

The results of applying these cuts are given in the panels of \autoref{fig:phot_qual_all} for each of the fields, with the CCHP field reproduced for ease of inter comparison (\autoref{fig:phot_qual_all}i). 
As anticipated from the shorter exposure times (Table~\ref{tab:datasets}), the quality parameters show an upturn in $\sigma_{F814W}$ and a flaring in sharp$_{F814W}$ at a brighter magnitude. 
The $\chi_{F814W}$ shows similar behavior for all panels. 
These restrictions were employed for all of the visualizations of the photometry presented in the main text as well as all of the analyses undertaken with these data.

\begin{table}
\centering
\footnotesize
\caption{\label{tab:calibration} Quantities Applied to the Photometry}
\begin{tabular}{l|ccccc}
\hline \hline
Quantity & M101\_{A}  & M101\_{B} & M101\_{C} & M101\_{D} & M101\_{E} \\
\hline \hline 
Observation Date    & 2015-09-09 & 2014-10-09 & 2016-02-19 & 2013-10-18 & 2014-02-15 \\
\hline
ZP F555W            & \nodata    & 25.721($\pm$0.02)    & 25.720($\pm$0.02)    & \nodata    & \nodata    \\
APInf F555W         & \nodata    & \apinffivefivefive ($\pm$0.02)& \apinffivefivefive($\pm$0.02) & \nodata & \nodata \\
ApCore$_{c1}$ F555W & \nodata    & $-$0.004($\pm$0.012) & 0.035($\pm$0.033) & \nodata & \nodata \\
ApCore$_{c2}$ F555W & \nodata    & $-$0.044($\pm$0.025) & $-$0.021($\pm$0.016) & \nodata & \nodata \\
\hline
ZP F606W            & 26.411($\pm$0.02)  & \nodata & \nodata & 26.413($\pm$0.02)  & 26.413($\pm$0.02)  \\
APInf F606W         & \apinfsixzerosix($\pm$0.02)  & \nodata & \nodata & \apinfsixzerosix($\pm$0.02)  & \apinfsixzerosix($\pm$0.02)  \\
ApCore$_{c1}$ F606W & 0.007($\pm$0.042) & \nodata & \nodata &  0.065($\pm$0.035) & 0.006($\pm$0.033) \\
ApCore$_{c2}$ F606W & $-$0.056($\pm$0.036) & \nodata & \nodata & 0.022($\pm$0.037) & $-$0.023($\pm$0.019)\\ 
\hline 
ZP F814W            &  25.523($\pm$0.02)   & 25.524($\pm$0.02)   &  25.523($\pm$0.02)  &  25.524($\pm$0.02)   &  25.524($\pm$0.02)   \\ 
APInf F814W         & \apinfeightonefour($\pm$0.02)  & \apinfeightonefour($\pm$0.02)  & \apinfeightonefour($\pm$0.02)  & \apinfeightonefour($\pm$0.02)  & \apinfeightonefour($\pm$0.02)  \\
ApCore$_{c1}$ F814W & 0.101($\pm$0.024) & 0.074($\pm$0.006) & 0.067($\pm$0.008) & 0.063($\pm$0.008) & 0.049($\pm$0.019) \\
ApCore$_{c2}$ F814W & 0.063($\pm$0.020) & 0.044($\pm$0.006) & 0.050($\pm$0.005) & 0.056($\pm$0.013) & 0.061($\pm$0.013)\\
\hline 
A$_{I,MW}$ [mag] & \IextinctioROUNDED ($\pm$\IextinctionerrROUNDED) & \aic($\pm$\eaic) & \aid($\pm$\eaid) & \aie($\pm$\eaie) & \aif($\pm$\eaif) \\
\hline 
m$_{\rm \sigma}$ [mag] & 22.25 & 21.50 & 21.75 & 21.50 & 21.50 \\
m$_{\rm sharp}$  [mag] & 23.75 & 22.50 & 22.50 & 22.50 & 22.50 \\
m$_{\rm \chi}$   [mag] & 23.00 & 23.00 & 23.00 & 23.00 & 23.00 \\ 
\hline \hline 
\end{tabular}

\end{table}

\subsection{Color-Magnitude Selection Box}\label{app:color_box} 

The color magnitude selection box (blue shading in \autoref{fig:our_trgb})  that were applied to the color-magnitude data before building the luminosity functions is defined as follows:
\begin{equation} \label{eq:cm-box}
\begin{split}
m_{F814W} \geq& -6.0~(m_{F814W}-m_{F606W} - 1.0) + 25.10 \\
m_{F814W} \leq& -6.0~(m_{F814W}-m_{F606W} - 1.6) + 25.10 
\end{split}
\end{equation}
The slope of -6.0 mag color$^{-1}$ was measured from high signal-to-noise photometry for Local Group galaxies used in the RR~Lyrae arm of the CCHP \citetalias[see][]{beaton_2016}. 
The color-width of 0.6 mag was set based on the width of the RGB in \autoref{fig:our_trgb}i. 
The data were also restricted to the color and magnitude ranges of the color-magnitude plot to avoid any confusion over differences between \autoref{fig:our_trgb}i and \autoref{fig:our_trgb}ii. 
\autoref{eq:cm-box} was transformed into the the F555W-F814W color-system following \autoref{eq:colortransform} to set the equivalent color-range. 

\subsection{Full Visualizations for Archival Fields} 
\responsetoref{ The full visualization akin to \autoref{fig:our_trgb_stable} is provided for each of the archival fields is provided in the panels of \autoref{fig:our_trgb_stable}. 
More specifically, the a-subpanels have the LF plotted for a range of smoothing widths (\sigmasmooth), the b-subpanels have the edge response ($\eta$) for that same range of smoothing widths, the c-panels are zooms of the normalized $\eta$ around $\eta_{max}$ for a subset of \sigmasmooth, and the d-panels demonstrate different means of quantifying the uncertainty on the determination of $m_{\eta_{max}}$. }

\begin{figure*} 
\centering
    \includegraphics[width=0.3\columnwidth]{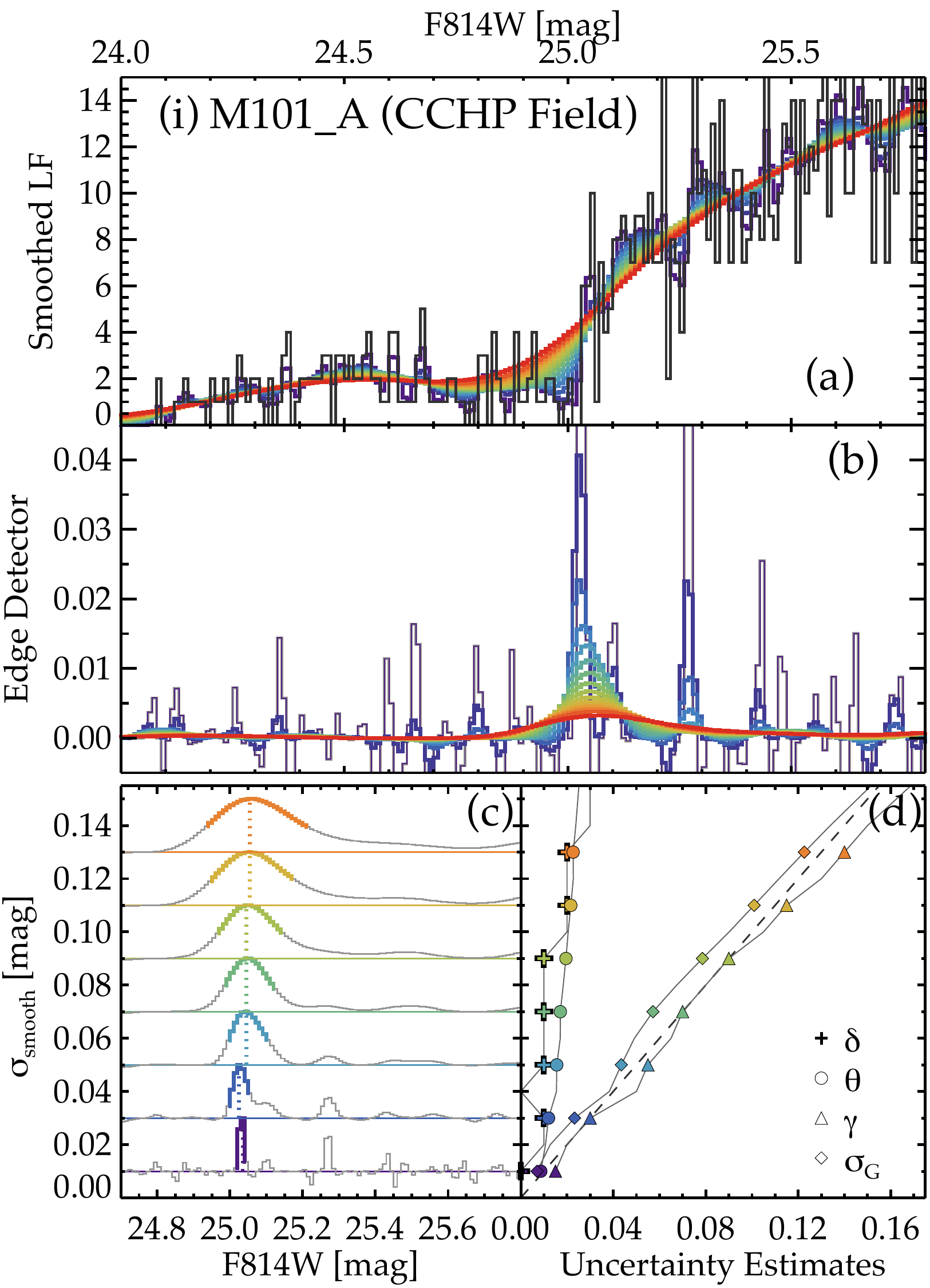}
    \includegraphics[width=0.3\columnwidth]{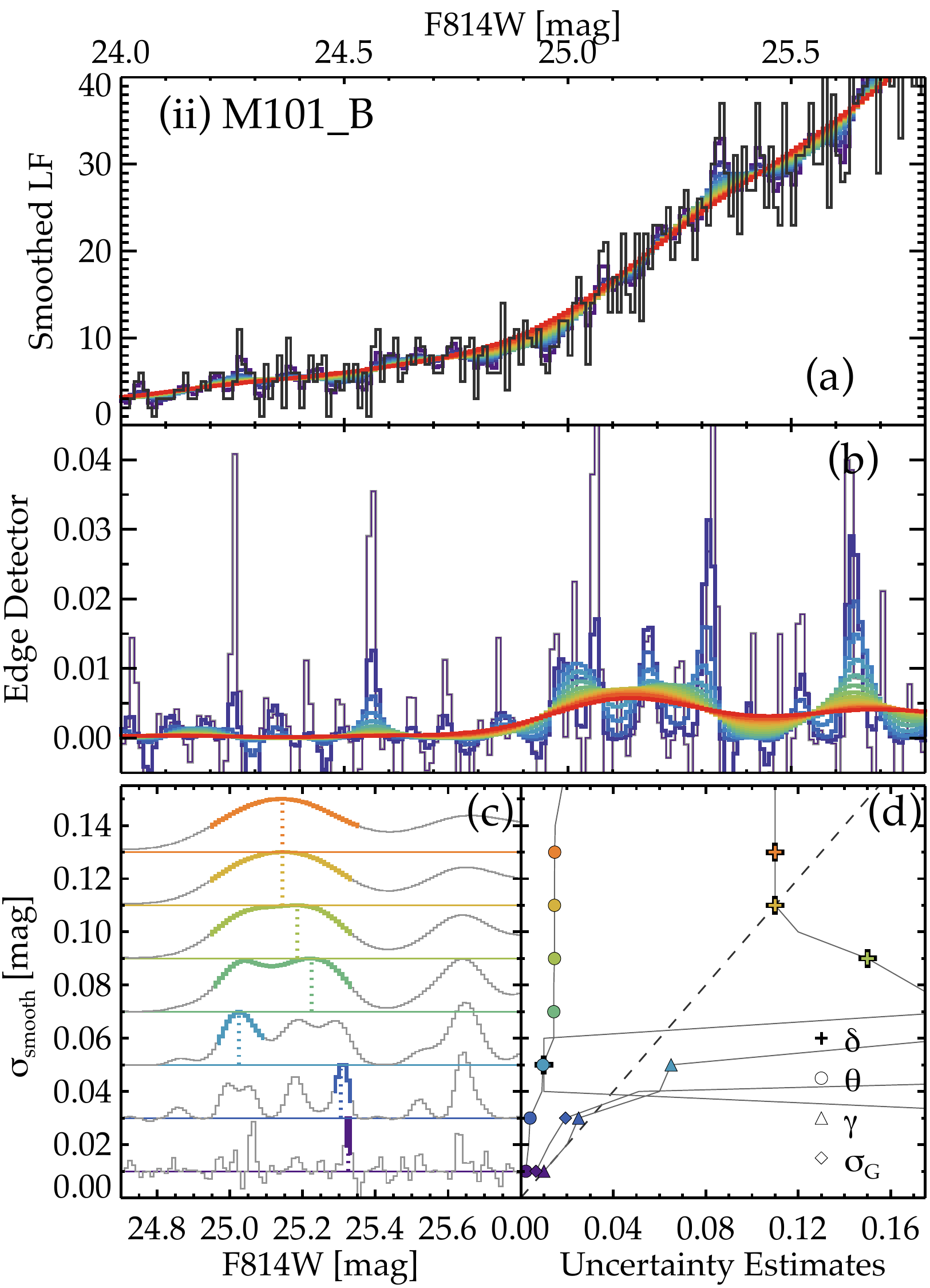}
 \\
    \includegraphics[width=0.3\columnwidth]{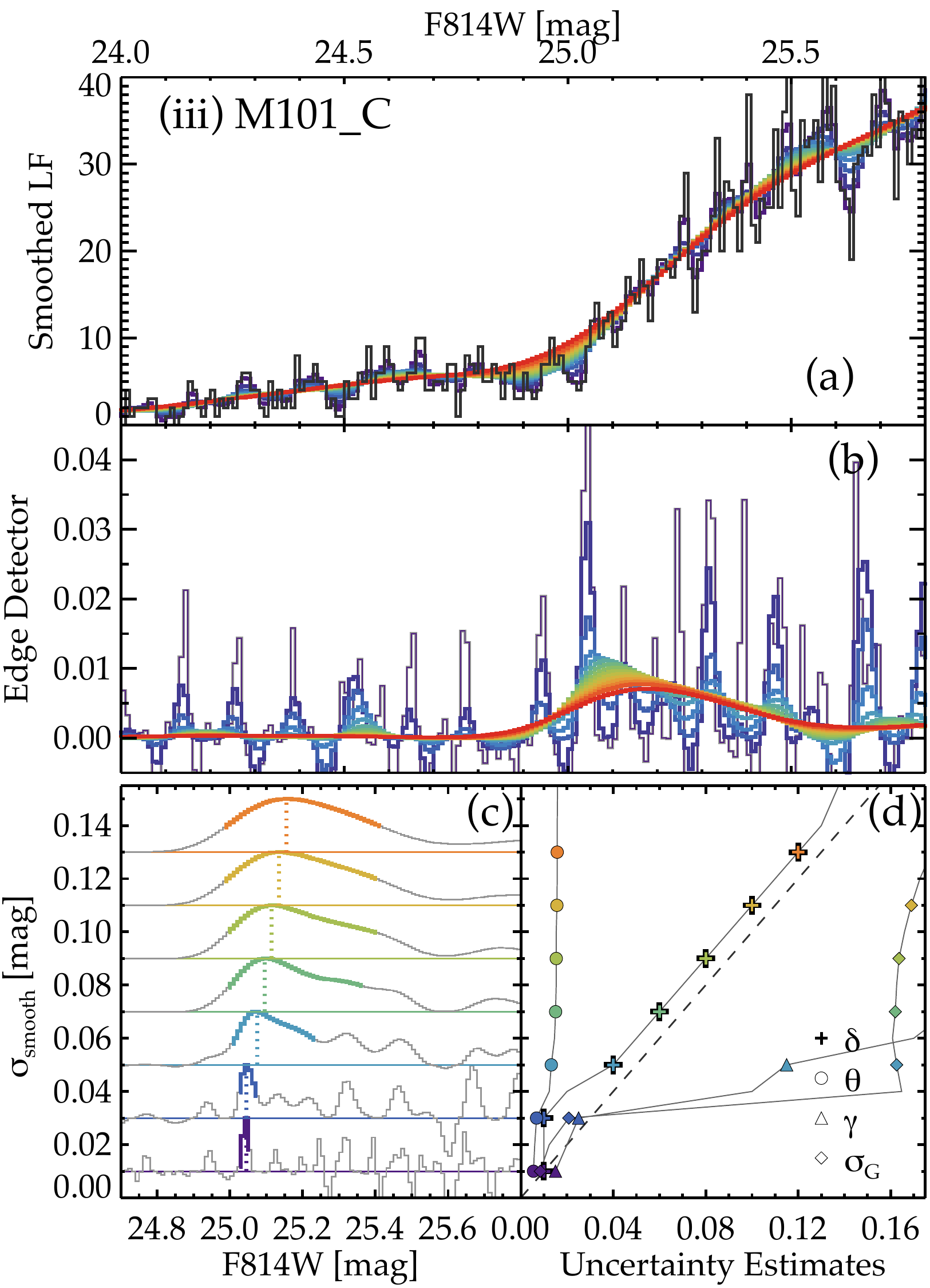} 
    \includegraphics[width=0.3\columnwidth]{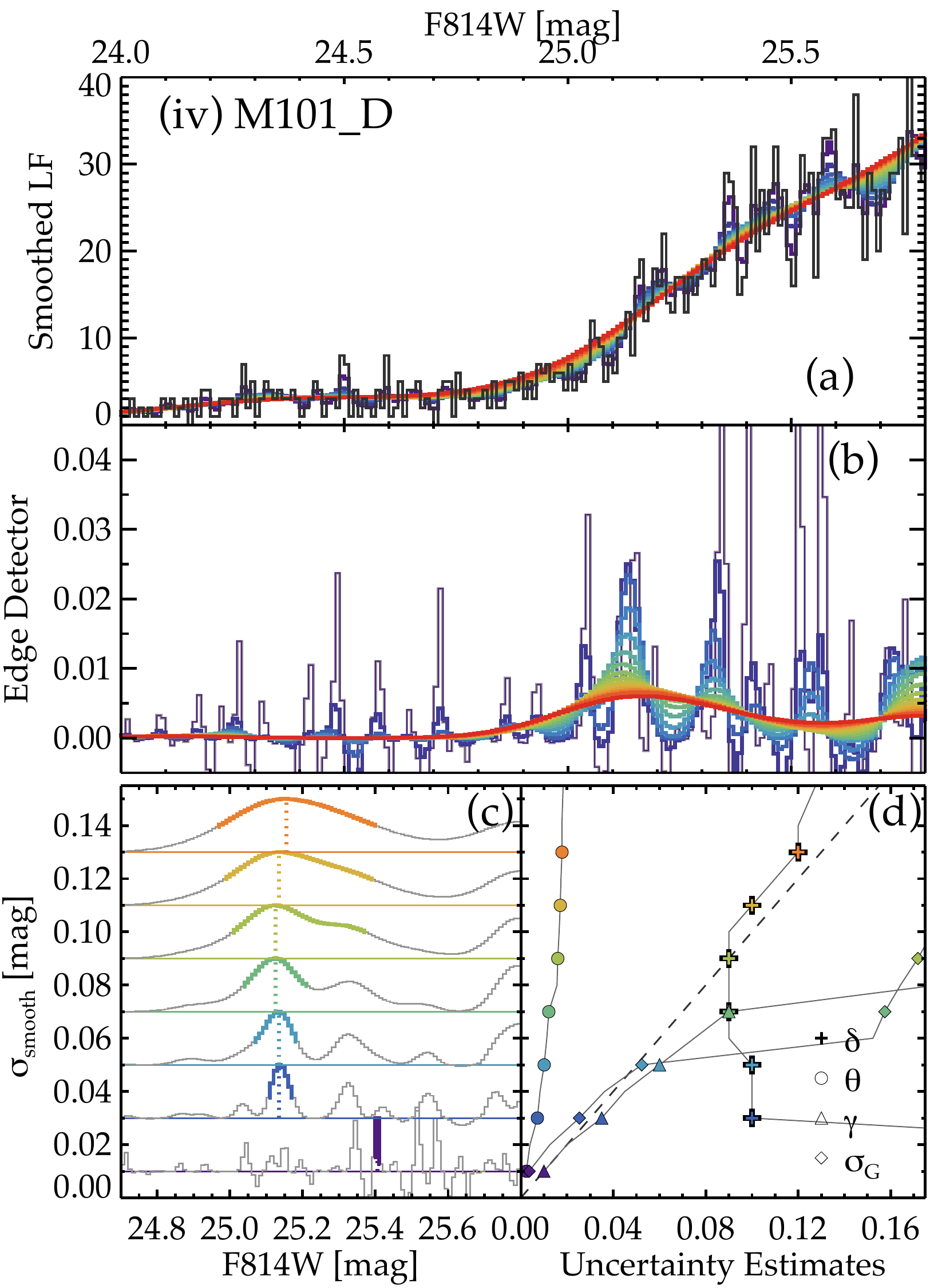}
    \includegraphics[width=0.3\columnwidth]{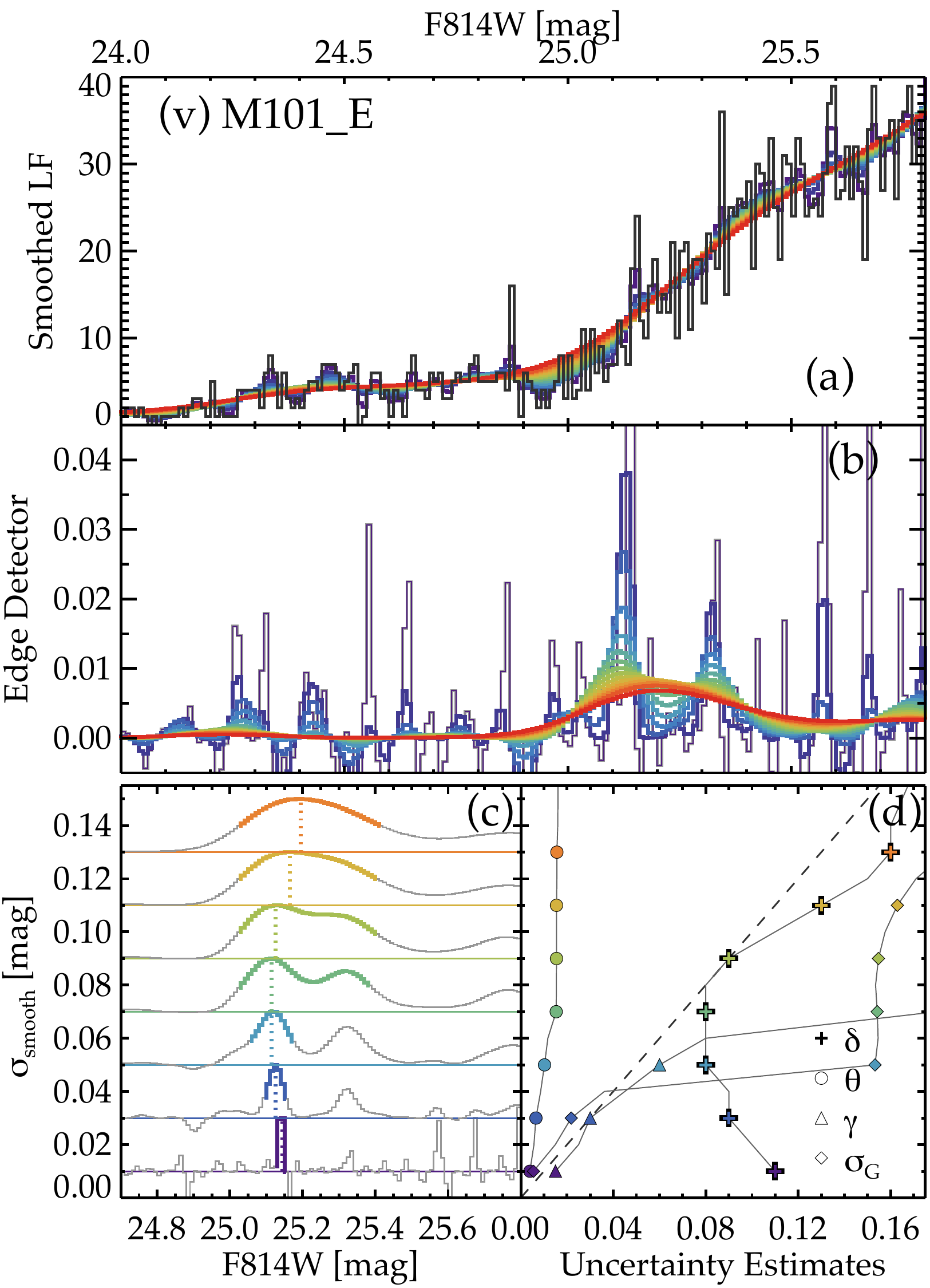}
\caption{\label{fig:smoothing_full} The impact of smoothing the luminosity function on TRGB measurements. Each panel is for an individual field, as follows: (i) M101\_A (CCHP Field), (ii) M101\_B, (iii) M101\_C, (iv) M101\_D, and (v) M101\_E. 
The sub-panels are identical to that of \autoref{fig:our_trgb_stable} and the panel for M101\_A is repeated for ease of comparison. 
While the LF (a-subpanels) do not show obvious differences, the edge-response ($\eta$) shows more complex behavior than in M101\_A (b-subpanels), which when normalized (c-subpanels) show that there is great ambiguity in the determination of $\eta_{max}$. The statistical metrics explored for M101\_A (d-subpanels) are complex and difficult to parse. We conclude that, unlike the CCHP field, the archival fields do not show a clear TRGB measurement.}
\end{figure*} 

\subsection{Measurement of TRGB in M101\_E}\label{app:jl17_comp}
The M101\_E field is the same image dataset used by \citetalias{janglee_2017}. To facilitate comparison between the \citetalias{janglee_2017} result and our own, we have performed a more detailed analysis of this field to the same level as for our CCHP field (M101\_A). 

\begin{figure} 
\centering
\includegraphics[width=0.5\columnwidth]{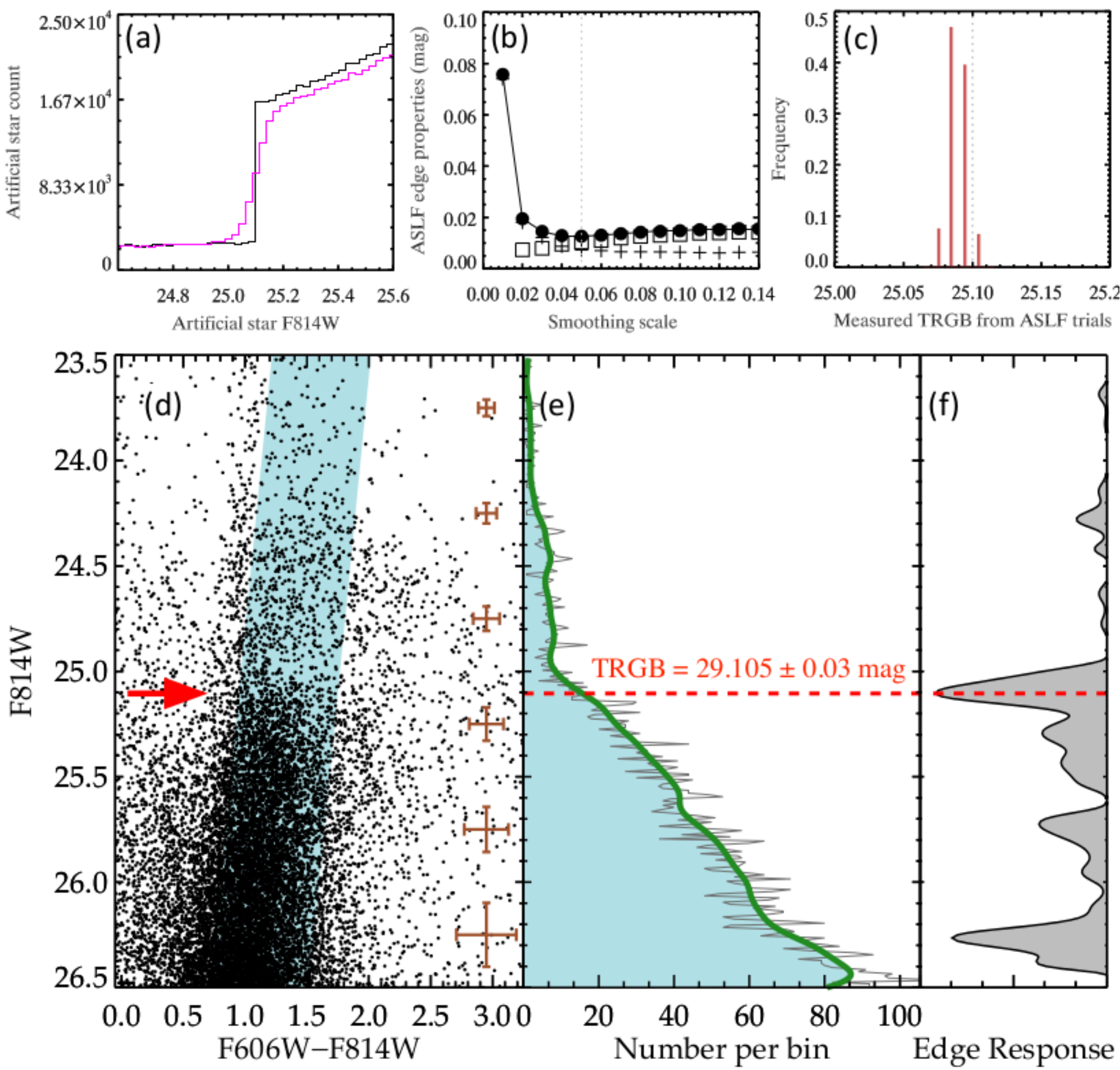}
\caption{\label{fig:aslf_results2}
Summary of the ASLF procedure to determine the optimal \sigmasmooth and its measurement uncertainties for the M101\_E field (a-c) and the TRGB measurement in the M101\_E field (d-f). 
(a) A comparison of the input (black) and output (purple) ASLF -- note that only the AGB and RGB are modeled here. 
(b) The systematic (open box), statistical (plus), and quadratic sum (filled circle) uncertainties determined via TRGB measurements of 10,000 realizations of the ASLF to match our source counts. 
(c) The distribution of TRGB measurements for the ``optimal'' \sigmasmooth of 0.05~mag. 
These uncertainties are smaller than those in M101\_A (\autoref{fig:aslf_results}) due to the larger number of sources. 
Due to the simplicity of our modeling (only AGB + RGB) and the complexity of the field, we suspect these uncertainties should only be considered a lower limit on the true uncertainties. 
(d) CMD and a color-cut (blue-shaded region) similar to that used in the CCHP field. 
(e) Raw LF binned at 0.01~mag (thin gray) and GLOESS smoothed LF using \sigmasmooth~=~0.05~mag (thick green).
(f) Edge detection response function with a maximum at \responsetoref{\tipf~$\pm$ \trgbobsvalstaterrROUNDED~mag}.} 
\end{figure}

Thus, identical procedures were followed as for the CCHP Field and the supporting figures are given here for completeness. 
The ASLF were constructed, inserted into the images, photometered, and analyzed. 
The panels of \autoref{fig:aslf_results2} give the steps of this process; more specifically, \autoref{fig:aslf_results2}a shows the input and output LF, \autoref{fig:aslf_results2}b summarizes the statistical, systematic and total uncertainties for each \sigmasmooth, and \autoref{fig:aslf_results2}c provides the distribution of TRGB measurements for the selected \sigmasmooth from which the final uncertainties are adopted. 
The resulting \sigmasmooth is 0.05~mag, identical to that of the CCHP Field and the uncertainties are similar, with 0.03~mag and 0.04~mag for the statistical and systematic uncertainties, respectively.

We note, however, that our ASLF analysis here is somewhat superficial given our knowledge of the field.
More specifically, we have only modeled, effectively, a single stellar population; whereas it is evident from the CMD that the M101\_E contains many different populations (e.g., stellar sequences of different ages and metallicities) that are superimposed as part of different galactic structures (with different internal extinctions that shift the apparent distance). 
With this in mind, we suspect that our estimated uncertainties on the TRGB detection should be taken as lower limits on the true uncertainties. 

The panels of \autoref{fig:aslf_results2} demonstrate the TRGB detection. 
The CMD is given in \autoref{fig:aslf_results2}i, with the blue shading giving a color-selection box that is broadened from that in \autoref{fig:our_trgb}i proportional to the photometric uncertainties. \autoref{fig:aslf_results2}ii gives the LF in both its raw and GLOESS smoothed form. \autoref{fig:aslf_results2}iii is the response function of the Sobel kernel showing a clear peak at \responsetoref{\tipf~$\pm$ \trgbobsvalstaterrROUNDED~mag, with the quoted uncertainty being the statistical uncertainty on the TRGB detection.} 

\subsection{Literature Distances to \gal} \label{app:litdistances}
\autoref{tab:distances} gives the individual values, uncertainties, references and notes for the distances to \gal that were used to construct \autoref{fig:distcomp}. 

\begin{table*} 
\centering
 \caption{\label{tab:distances} Literature Distances to M\,101}
\begin{tabular}{ l c l } 
 \hline \hline
 \multicolumn{3}{l}{Measurement of the {\sc TRGB}} \\
 Study & TRGB Magnitude (F814W) & Notes \\
 \hline  
 \citet{sakai_2004} (\citetalias{sakai_2004}) & 25.40 $\pm$ 0.04 & \\
 \citet{rizzi_2007} (\citetalias{rizzi_2007}) & 25.31 $\pm$ 0.08 & \\
 \citet{shappee_2011} (\citetalias{shappee_2011}) & 25.00 $\pm$ 0.06 & \\
 \citet{lee_2012} (\citetalias{lee_2012}) & 25.28 $\pm$ 0.01 & Concerns over photometric zero point. \\
 \citet{Tikhonov_2015} (\citetalias{Tikhonov_2015}) & 25.05 & no uncertainties given \\
                       & 25.10 & Pointing 1; no uncertainties given \\ 
                       & 25.11 & Field M101\_D; no uncertainties given \\
 \citet{janglee_2017b} (\citetalias{janglee_2017}) & 25.16 $\pm$ 0.04 & \\ 
 \responsetoref{EDD \citep{jacobs_2009}} & \responsetoref{25.08} +0.02 -0.03 & \responsetoref{Independent reduction of CCHP Pointing} \\  
 CCHP (This Work) & 25.04 $\pm$ 0.03 & \\ 
  \hline \hline 
\multicolumn{3}{l}{Distances determined via the {\sc Leavitt Law}} \\
 Study & Distance Modulus & Notes \\
  \hline  
 \citet{kelson_1996}    & 29.24 $\pm$ 0.10  & \\ 
 \citet{stetson_1998}   & 29.05 $\pm$ 0.14 & \\
                        & 29.21 $\pm$ 0.17 & \\
 \citet{Kennicutt_1998} & 29.20 $\pm$ 0.07 & \\
                        & 29.34 $\pm$ 0.08 & \\
                        & 29.39 $\pm$ 0.07 & \\
 \citet{ferrarese_2000} & 29.34 $\pm$ 0.10 & \\
 \citet{macri_2001}     & 29.04 $\pm$ 0.08 & NIR using \hst~+~NICMOS \\
                        & 29.77 $\pm$ 0.09 & \\
                        & 29.58 $\pm$ 0.09 & \\
                        & 29.45 $\pm$ 0.08 & \\
                        & 29.37 $\pm$ 0.01 & \\
 \citet{newman_2001}    & 29.06 $\pm$ 0.11 & \\
                        & 29.16 $\pm$ 0.09 & \\
 \citet{Willick_2001}   & 29.21 $\pm$ 0.08 & \\
 \citet{freedman_2001} (KP) & 29.13 $\pm$ 0.11 & Final KP Result \\
 \citet{paturel_2002}   & 29.23 $\pm$ 0.07 & $\mu_{LMC}$ = 18.37 mag\\
                        & 29.26 $\pm$ 0.15 & \\
                        & 29.30 $\pm$ 0.07 & \\
 \citet{sakai_2004}     & 29.14 $\pm$ 0.09 & \\
                        & 29.24 $\pm$ 0.08 & \\
 \citet{saha_2006}      & 29.18 $\pm$ 0.08 & $\mu_{LMC}$ = 18.54 mag\\
 \citet{shappee_2011} (SS11) & 29.04 $\pm$ 0.04 & $\mu_{LMC}$ = 18.41 mag \\
 \citet{mager_2013} & 28.96 $\pm$ 0.11 & $\mu_{LMC}$ = 18.48 mag\\
 \citet{Tully_2013} & 29.21 $\pm$ 0.06 & Cosmicflows-2 Compilation \\
 \citet{nataf_2015} & 29.20 $\pm$ 0.03 & Uses SS11, but A$_{I}$/E(V-I)=1.1450; $\Delta \mu_{LMC}$=10.72 $\pm$ 0.03 mag \\
 \citet{riess_2016} & 29.14 $\pm$ 0.05 & \\ 
 \hline \hline
\end{tabular} 
\end{table*} 

\end{document}